\documentclass[twocolumn,preprintnumbers,amsmath,amssymb,superscriptaddress,nofootinbib,longbibliography]{revtex4-1}

\usepackage{graphicx}
\usepackage{bm}
\usepackage{dsfont}
\usepackage[usenames,dvipsnames]{xcolor}
\usepackage{pstricks}
\usepackage[tight]{subfigure}
\usepackage{verbatim}
\usepackage{units}
\usepackage{multirow}
\usepackage{enumitem}
\usepackage{mathrsfs}
\usepackage{leftidx}
\usepackage{xspace}
\usepackage{braket}
\usepackage{bbm}
\usepackage{amsfonts,amssymb,amsmath}

\usepackage[a4paper]{hyperref}
\hypersetup{colorlinks=true,linktoc=all,linkcolor=blue,breaklinks=true,citecolor=blue,urlcolor=blue}
\usepackage[english]{babel}
\newcommand{\rfs}{resource fluctuations\xspace}
 \newcommand{\Scl}{S_\text{res,cl}}
 \newcommand{\Stot}{S_\text{res}}
\newcommand{\kB}{k_\text{B}}
\newcommand{\Isigma}{I^\Sigma}

\newcommand{\Ienergy}{I^\mathrm{E}}

\usepackage[normalem]{ulem}

\usepackage{hyperref}

\AtBeginDocument{%
    \newwrite\bibnotes
    \def\bibnotesext{Notes.bib}
    \immediate\openout\bibnotes=\jobname\bibnotesext
    \immediate\write\bibnotes{@CONTROL{REVTEX41Control}}
    \immediate\write\bibnotes{@CONTROL{%
    apsrev41Control,author="08",editor="1",pages="1",title="0",year="1"}}
     \if@filesw
     \immediate\write\@auxout{\string\citation{apsrev41Control}}%
    \fi
}%

\begin{document}
	
	
    \title{Constraints between entropy production and its fluctuations in nonthermal engines}
	
	\author{Matteo Acciai}
    \thanks{These authors contributed equally to this work.}
	\affiliation{Department of Microtechnology and Nanoscience (MC2), Chalmers University of Technology, S-412 96 G\"oteborg, Sweden\looseness=-1}

    \author{Ludovico Tesser}
    \thanks{These authors contributed equally to this work.}
	\affiliation{Department of Microtechnology and Nanoscience (MC2), Chalmers University of Technology, S-412 96 G\"oteborg, Sweden\looseness=-1}

	\author{Jakob Eriksson}
	\affiliation{Department of Microtechnology and Nanoscience (MC2), Chalmers University of Technology, S-412 96 G\"oteborg, Sweden\looseness=-1}
	
	\author{Rafael S\'anchez}
	\affiliation{Departamento de Física Te\'orica de la Materia Condensada, Condensed Matter Physics Center (IFIMAC), and Instituto Nicol\'as Cabrera, Universidad Aut\'onoma de Madrid, 28049 Madrid, Spain\looseness=-1}
	
	\author{Robert S. Whitney}
	\affiliation{Universit\'e Grenoble Alpes, CNRS, LPMMC, 38000 Grenoble, France}

	\author{Janine Splettstoesser}
	\affiliation{Department of Microtechnology and Nanoscience (MC2), Chalmers University of Technology, S-412 96 G\"oteborg, Sweden\looseness=-1}
	
	\date{\today}
	
	\begin{abstract}
We analyze a mesoscopic conductor autonomously performing a thermodynamically useful task, 
such as cooling or producing electrical power, 
in a part of the system---the working substance---by exploiting 
another terminal or set of terminals---the resource---that contains a stationary nonthermal (nonequilibrium) distribution.
Thanks to the nonthermal properties of the resource, \textit{on average} no exchange of particles or energy with the working substance is required to fulfill the task. 
This resembles
the action of a \textit{demon}, as long as only average quantities are considered. 
Here, we go beyond a description based on average currents and investigate the role of fluctuations in such a system. 
We show that a minimum level of entropy fluctuations in the system is necessary, whenever one is exploiting a certain entropy production in the resource terminal to perform a useful task in the working substance.
For concrete implementations of the demonic nonthermal engine in three- and four-terminal electronic conductors in the quantum Hall regime, we compare the resource fluctuations to the entropy production in the resource and to the useful engine output (produced power or cooling power).
	\end{abstract}

	\maketitle

\section{Introduction}
The rapidly evolving research field of quantum thermodynamics is constantly pushing the boundaries of our understanding of thermodynamic processes at the nanoscale~\cite{QuThermoBook,Strasberg2022Book}. In contrast to macroscopic systems, which are typically well described by average quantities, small-scale devices are very much influenced by fluctuations~\cite{Esposito2009Dec,Seifert2008Aug} and often possess quantum features, requiring the standard thermodynamic laws to be modified and complemented by additional concepts, for example borrowed from information theory. Besides the fundamental challenge of establishing a thermodynamically consistent description of nanoscale systems, understanding their behavior leads to potential practical applications. A prominent example is that of heat management and energy conversion in nanoelectronics~\cite{Benenti2017Jun}, with an ever increasing relevance due to the miniaturization opportunities offered by technological developments. 
\begin{figure}[b]
\includegraphics[width=\columnwidth]{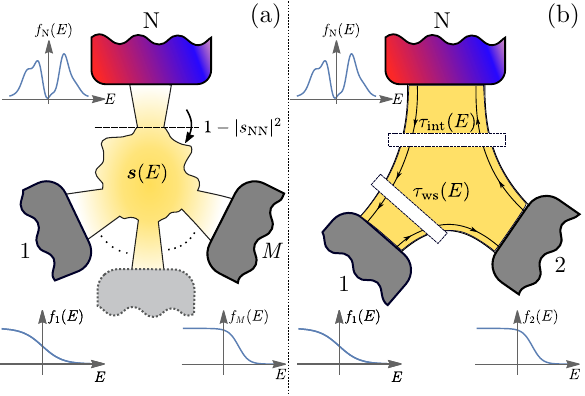}
\caption{\label{fig:setup_3T}
(a) Generic multi-terminal electronic conductor fed by a nonthermal resource (N) with distribution $0\leq f_\text{N}(E)\leq 1$. The properties of the coherent conductor are encoded in the scattering matrix $\boldsymbol{s}(E)$. In particular, the resource region (above the horizontal dashed line) is connected to the working substance (below the dashed line) by the transmission probability $1-|s_\text{NN}(E)|^2$. (b) Minimal setup exploiting a nonthermal resource in a
a 3-terminal quantum Hall device. The resource is connected to the working substance by the interface transmission probability $\tau_\text{int}(E)$ and power production is enabled by the energy-dependent transmission $\tau_\mathrm{ws}(E)$.
}
\end{figure}

Among the exciting features of small-scale systems is the possibility to access non-conventional resources that can be used to power nanoscale engines, in addition to heat and work. 
It is for instance interesting to understand the role of quantum coherences~\cite{Scully2003Feb,Francica2017Mar,Niedenzu2018Jan,Ghosh2019Mar}, as well as that of information~\cite{Bennett1982Dec,Berut2012Mar,Parrondo2015Feb} {or of active matter~\cite{Krishnamurthy2016Dec,Holubec2020Nov,Pietzonka2019Nov}.} 
Here, we focus on another option, namely to exploit a {terminal that supplies a steady-state fermionic resource that is nonthermal (nonequilibrium)~\cite{Abah2014May,Tesser2023Apr}. It hence does} not possess a well-defined temperature or potential, as depicted in Fig.~\ref{fig:setup_3T}(a). 
While this contradicts the standard notion of a bath or a reservoir, it is a quite common situation in nanoscale systems {that might interact with different environments and that can have themalization lengths exceeding} the system size~\cite{Benenti2017Jun}. Also time-dependent nonequilibrium states can be available~\cite{Strasberg2017Apr,Konopik2020Nov,Ryu2022May}. Moreover, it is likely that such nonthermal resources are present in nanoscale systems, as they may originate as a ``waste'' byproduct of other processes of interest. Understanding how such nonthermal distributions can be harvested is hence expected to be relevant from a practical point of view.

In previous works~\cite{Whitney2016Jan,Sanchez_2019}, some of us have introduced the concept of a nonequilibrium demon (N-demon). 
Such a system is able to perform a useful task in a steady-state operation by solely relying on a nonthermal (nonequilibrium) resource, in the absence of \textit{average} heat or work consumption, namely neither average energy nor average particle currents flow between the resource region and the working substance, see Fig.~\ref{fig:setup_3T}. 
This resembles a Maxwell demon~\cite{Esposito2012Aug,Strasberg2013Jan,Koski2014Sep,Koski2015Dec,chida:2017,Masuyama2018Mar,Sanchez2019Oct,Freitas_2021,Bozkurt2022Dec,Saha2022Jul}---hence the choice of the name N-demon~\cite{Whitney2023Apr}---but the working principle is very different as it does not rely on any type of feedback and is ``demonic" only when focusing on average quantities. This has triggered a number of related works, where various properties of systems relying on nonequilibrium resources have been analyzed~\cite{Deghi2020Jul,Ciliberto2020Nov,Hajiloo_2020,Lu2021Feb}.
In particular, it has been shown that a performance characterization of an N-demon~\cite{Deghi2020Jul,Hajiloo_2020} can be achieved by introducing well-behaved free-energy efficiencies~\cite{Manzano2020Dec,Hajiloo_2020,Tesser2023Apr} taking into account that a nonthermal resource is exploited.

The absence of feedback mechanisms in an N-demon suggests that fluctuations play a fundamental role for the operation of engines exploiting nonthermal resources~\cite{Freitas2021Mar}. 
However, while noise in the absence of charge or heat currents has recently attracted a lot of interest in systems with standard thermal reservoirs~\cite{Lumbroso2018Oct,Larocque2020Sep,Eriksson_2021,Tesser2022Oct}, previous studies on nonthermal engines have addressed average quantities only. 
The characterization of fluctuations of nonthermal resources  for the operation of nanoscale engines in general, and for N-demons more specifically, is the gap we aim to fill with the present paper.
More concretely, we quantify to what extent the fluctuations associated with temporary exchanges of particles and energy between the resource and the working substance relate to the performance of this autonomous (steady-state) device. 
In this paper, we identify that the entropy reduction in the working substance, and hence the entropy production in the resource region, requires a minimum level of fluctuations. We find that it is a combination of entropy and particle fluctuations that sets the constraint.
This key result reads
\begin{equation}
S_{\text{NN,cl}}^{\Sigma} {+} \frac{k_\text{B}^2}{4}S_{\text{NN,cl}}^{} \geq 2k_\text{B}\Isigma_{\rm N}\geq -2k_\text{B}\Isigma_{\rm ws},
\label{eq:result_intro}
\end{equation}
where $\Isigma_\text{N}$ is the entropy production in the nonthermal resource, $-\Isigma_\text{ws}$ is the entropy reduction in the working substance (quantifying the useful output){, where both the resource and working substance contain one or more fermionic steady-state contacts}. 
The terms on the left-hand side characterize the entropy and particle fluctuations in the nonthermal resource (see Sec.~\ref{sec:generalConstraints} for definitions and details of the introduced quantities).
We refer to this sum as the classical part of the \emph{\rfs} in the following. For systems with thermal contacts only, this results in constraints involving heat and particle fluctuations. In this paper, we derive Eq.~\eqref{eq:result_intro} and illustrate its implications with specific examples.
More specifically, we apply our findings, that are valid for arbitrary noninteracting multi-terminal conductors with possibly nonthermal resources, to 3- or 4-terminal N-demon systems realized in the quantum Hall regime. In conductors in the quantum Hall regime, the time-reversal symmetry breaking in chiral edge states~\cite{Buttiker1988Nov} allows for particularly simple implementations of quantum conductors for energy conversion~\cite{Sanchez2015Apr} and for the inspection of current-current correlations~\cite{texier:2000}. 
In experimental implementations of such systems, detection of nonthermal distributions~\cite{Altimiras2010Jan}, of heat flow~\cite{Jezouin2013Nov}, and of noise~\cite{Sivre2019Dec} have been demonstrated, making them relevant for our purpose.
We identify situations in which the noise is particularly small, approaching the discovered bound, Eq.~\eqref{eq:result_intro}. For specific, experimentally relevant implementations of N-demons in quantum Hall conductors, namely where either the nonthermal distribution allows one to get close to the identified bound or where the nonthermal distribution is engineered by a coherent mixing of thermal resources, we analyze the \rfs and how they are related to the maximum power or cooling power that can be achieved in the device.

This paper is structured as follows: We introduce the model and the employed scattering approach for all analyzed quantities, including the entropy currents and their fluctuations, in Sec.~\ref{sec:model}. We then demonstrate that they are related to each other in a general way by constraints on entropy fluctuations compared to entropy flow in Sec.~\ref{sec:generalConstraints}. In Sec.~\ref{sec:minimal-model}, we provide the shape that these constraints take for a three-terminal quantum Hall setup and analyze their implications. Finally, results for \rfs in a specific 4-terminal setup, where the nonthermal distribution is engineered, are presented in Sec.~\ref{sec:composite}. The Appendices contain analytical expressions in limiting regimes as well as derivations and results for the four-terminal setup, where the nonthermal distribution is injected from a resource consisting of a \textit{set} of thermal contacts, complementing the ones in the main paper.

\section{Model and thermodynamic transport quantities}\label{sec:model}

\subsection{Multi-terminal conductor}
We consider a generic multi-terminal coherent electronic conductor, as sketched in Fig.~\ref{fig:setup_3T}(a). 
The upper terminal, denoted by N, is in a nonthermal state and provides the resource exploited by the working substance---a coherent conductor with $M$ standard thermal reservoirs, denoted by $1, 2,\dots M$. 
The task of the device is to exploit the resource provided by the nonthermal terminal to produce in a steady-state operation a useful output in the working substance, such as extracting electrical power between any two contacts or cooling one of the contacts by transporting heat into a hotter contact. 
In general, the device does something useful whenever the entropy of the working substance is reduced, at the expense of an increase in the entropy of the resource region\footnote{This entropy production in the resource region also corresponds to the minimum cost required for maintaining the state of the resource.}. 
In this Section and in Sec.~\ref{sec:generalConstraints}, we will keep the treatment as general as possible, while a minimal model based on a three-terminal device, sketched in Fig.~\ref{fig:setup_3T}(b), will be presented in Sec.~\ref{sec:minimal-model}.

The theoretical framework we rely on is a scattering matrix approach~\cite{Blanter2000,moskalets-book},  valid for coherent conductors in which many-body interactions are negligibly small.
We emphasize that, interactions being absent, we can exclude that any autonomous feedback mechanisms~\cite{Strasberg2013Jan,Koski2014Sep,Whitney2016Jan,Sanchez2019Oct} play a role in our device, so that any ``demonic'' effects, appearing in the absence of average charge energy flow between resource and working substance, can safely be attributed to the presence of the nonthermal resource. 

The energy-dependent scattering matrix $\boldsymbol{s}(E)$ describes the conductor's properties, where the matrix element $s_{\alpha\beta}(E)$ gives the probability amplitude that an electron emitted at energy $E$ by terminal $\beta$ is transmitted into terminal $\alpha$. The energy dependence of the scattering matrix is typically tunable by gates in mesoscopic electronic conductors. The terminals are characterized by distribution functions denoted by $f_\alpha(E)$, with $\alpha=1,2,\dots M,\text{N}$. For a standard equilibrium contact, we have a Fermi function
\begin{equation}
    f_\alpha(E)=\frac{1}{1+\exp[\beta_\alpha(E-\mu_\alpha)]},\quad \alpha\neq\text{N},
\end{equation}
with inverse temperature $\beta_\alpha=1/k_\text{B}T_\alpha$ and electrochemical potential $\mu_\alpha$. In contrast, $f_\text{N}(E)$ is a generic function describing the nonthermal terminal N. It is only constrained by $0\le f_\text{N}(E)\le 1$, as it represents the occupation probability of an electronic (fermionic) contact.

\subsection{Charge, heat and entropy currents and fluctuations}\label{sec:observables}
The average, steady-state currents flowing into terminal $\alpha$ {can be found as the expectation values of the fluctuating current operators as provided in Appendix~\ref{app_entropyoperator} and} are given by~\cite{Blanter2000,moskalets-book}
\begin{equation}
    I^X_\alpha=\braket{\hat{I}_\alpha^{X}}=\frac{1}{h}\int dE\,x_\alpha\sum_{\beta}|s_{\alpha\beta}(E)|^2[f_\beta(E)-f_\alpha(E)]\,.\label{eq:currents}
\end{equation}
Here, we use the abbreviation $X$ in order to indicate different types of currents, with $(X,x_\alpha)\in\left\{(\emptyset,1),(\mathrm{E},E),(\Sigma,\kB\sigma_\alpha)\right\}$. This includes the particle current  $I_\alpha^{}$ and the energy current $\Ienergy_\alpha$. The average heat current $J_\alpha$, for $\alpha\neq\text{N}$, is consequently given by
\begin{equation}
    J_\alpha=\Ienergy_\alpha-\mu_\alpha I_\alpha\,.
\end{equation}
In addition to charge, energy, and heat currents, we are interested in the entropy production in a given terminal. In this steady-state coherent conductor, this is given by the entropy current into contact $\alpha$, namely $\Isigma_\alpha$, 
with 
{
\begin{equation}\label{eq:entropyfactor}
\sigma_\alpha(E) \equiv -\log\left[\frac{f_\alpha(E)}{1-f_\alpha(E)}\right],
\end{equation}
}
see Ref.~\cite{Deghi2020Jul} or Appendix~\ref{app_entropyoperator} for different derivations of the entropy current. This entropy current $\Isigma_\alpha$ properly reduces to Clausius' relation 
\begin{equation}
    \Isigma_\alpha\rightarrow I_\alpha^{\Sigma,\mathrm{thermal}}=\frac{J_\alpha}{T_\alpha}\,,
    \label{eq:clausius}
\end{equation}
valid for thermal contacts, when $f_\alpha$ is a Fermi function, i.e., $\alpha\neq \mathrm{N}$. {Moreover, it is a thermodynamically consistent quantity, as it satisfies the second law, $\sum_\alpha \Isigma_\alpha\ge 0$, even when all contacts have nonthermal occupation probabilities~\cite{Whitney2023Apr}. This formulation for the entropy production requires the nonthermal contact to be a large, macroscopic electronic system without any classical or quantum correlations, referred to as ``nonequilibrium incoherent reservoir'' by the authors of Ref.~\cite{Deghi2020Jul}.}

{We mention that a different formulation of entropy production has been introduced in Ref.~\cite{Bruch2018Mar}, in situations possibly involving time-dependent drivings. There, however, the authors are concerned with the entropy production associated with the scattering states \emph{in the coherent leads}, as opposed to the one referring to the \emph{macroscopic contacts} that we are considering here. The difference is that by looking at the outgoing states in the leads only, any equilibration process happening when the electrons enter the macroscopic contact is not taken into account.}

We refer to the multi-terminal system introduced above as an N-demon and to the resource region as ``demonic", whenever the system obeys the strict \emph{demon conditions} on the currents
\begin{equation}
I_\text{N}=\Ienergy_\text{N}=0\,.
\label{eq:demon-conditions}
\end{equation}
This enforces that no exchange of particles or energy between the working substance and the resource region happens \textit{on average}. 
In this case, any useful output generated in the working substance is obtained without the consumption of an average energy resource and the resource is characterized by entropy flows or by free energies~\cite{Hajiloo_2020}. 
Even though Eq.~\eqref{eq:demon-conditions} rules out any net transfer of particles and energy, these quantities can still fluctuate. We will show in this paper that such fluctuations must be present in order for the nonthermal resource to produce a useful output under the strict demon conditions.

The zero-frequency noise related to any of the currents $I_\alpha^{X}$ considered here is defined by the correlator
\begin{equation}
    S_{\alpha\beta}^{X}=\int_{-\infty}^{+\infty}dt\Braket{\{\delta \hat{I}_\alpha^{X}(t),\delta \hat{I}_\beta^{X}(0)\}},
\label{eq:noise-def}
\end{equation}
where 
$\delta\hat{I}_\alpha^{X}=\hat{I}_\alpha^{X}-I_\alpha^{X}$ are the fluctuations of the current with respect to its average, steady-state value{, see Appendix~\ref{app_entropyoperator} for the definitions of the relevant fluctuating current operators}.
For the general case sketched in Fig.~\ref{fig:setup_3T}(a), these quantities evaluate to 
$S_{\alpha\beta}^{X}=S_{\alpha\beta,\text{cl}}^{X}+S_{\alpha\beta,\text{qu}}^{X}$, with~\cite{moskalets-book,nazarov_blanter_2009,Blanter2000}
\begin{subequations}
\begin{gather}
\begin{split}
    S_{\alpha\beta,\text{cl}}^{X}&\equiv\frac{2}{h}\int dE\,x_\alpha x_\beta
    \Big\{-F_{\alpha\alpha}|s_{\beta\alpha}|^2-F_{\beta\beta}|s_{\alpha\beta}|^2\\
    &\quad+\delta_{\alpha\beta}\sum_{\gamma} |s_{\alpha\gamma}|^2[F_{\alpha\gamma}+F_{\gamma\alpha}]\Big\}\,,
\end{split}\\
\begin{split}
    S_{\alpha\beta,\text{qu}}^{X}\equiv -\frac{2}{h}\int dE\,x_\alpha x_\beta&\Re\Big\{ \Big[\sum_{\gamma}(f_\alpha-f_\gamma) s^*_{\alpha\gamma}s_{\beta\gamma}\Big]\\
    &\times \Big[\sum_{\gamma}(f_\beta-f_\gamma) s_{\alpha\gamma}s_{\beta\gamma}^*\Big]\Big\}.
\end{split}
\end{gather}
\label{eq:noise-general}
\end{subequations}
Here, the total noise has been split into classical-like and quantum contributions. The classical-like part contains all the linear terms in the scattering probabilities $|s_{\alpha\beta}|^2$ and describes one-particle transfers across the conductor. 
It can be seen as a generalization of a fully classical expression with the addition of Pauli-blocking factors $F_{\alpha\beta}\equiv f_\alpha(1-f_\beta)$ implementing the exclusion principle~\cite{Brandner2018PRL,Liu2019Jun,Kirchberg2022Nov} (further quantum effects can still influence the form of $|s_{\alpha\beta}|^2$). 
The quantum part, in contrast, involves correlated two-particle processes. Note that the division of noise into separate meaningful contributions is obviously not unique~\cite{Buttiker1992Nov,Blanter2000};  for the situation studied here, the separation into quantum and classical contributions turns out to be favourable.
In Eq.~\eqref{eq:noise-general}, we have omitted for convenience the energy dependence of the scattering matrix elements $s_{\alpha\beta}(E)$ and of the distribution functions $f_\alpha(E)$. 

In what follows, a relevant role will be played by the entropy fluctuations
$S_{\alpha\beta}^\Sigma$ 
for generic, thermal or nonthermal occupation probabilities $f_\alpha$ and $f_\beta$. For entropy fluctuations between terminals $\alpha$ and $\beta$ with \textit{thermal} distributions, these fluctuations are nothing but the correlation functions $\braket{\delta\hat{J}_\alpha\delta\hat{J}_\beta}/(T_\alpha T_\beta)$ of the heat currents, divided by the corresponding temperatures of terminals $\alpha$ and $\beta$.
{We stress that a straightforward connection between the entropy and heat can be obtained in the thermal case only. When contacts with nonthermal occupation probabilities are involved, inferring the entropy production necessarily relies on the knowledge of the nonthermal distributions themselves. While this can be a challenging task, it has been shown that such nonequilibrium distributions can be experimentally accessed via spectroscopy techniques~\cite{Altimiras2010Jan}.}

\section{General bound on local entropy fluctuations}\label{sec:generalConstraints}
We are now in a position to answer the question raised in the introduction: 
What is the minimum amount of fluctuations of the resource currents (possibly with zero average) required for the production of a useful output in the working substance?

We here identify the entropy production and its fluctuations as well as the particle current fluctuations, as the relevant quantities which constrain each other. Specifically, for the autocorrelations of the entropy currents in any terminal $\alpha$, $S_{\alpha\alpha}^\Sigma = S_{\alpha\alpha, \text{cl}}^\Sigma + S_{\alpha\alpha, \text{qu}}^\Sigma$, one has for the classical and quantum parts
\begin{subequations}
\begin{align}
\begin{split}
    S_{\alpha\alpha,\text{cl}}^{\Sigma}=&\frac{2k_\text{B}^2}{h}\int dE\,\left[\sigma_\alpha\right]^2
    \sum_{\gamma\neq\alpha}|s_{\alpha\gamma}|^2(F_{\alpha\gamma}+F_{\gamma\alpha}),
    \end{split}
    \label{eq:noise_classical}\\
     S_{\alpha\alpha,\text{qu}}^{\Sigma}=&-\frac{2k_\text{B}^2}{h}\int dE\,\Big[\sigma_\alpha\sum_{\gamma\neq\alpha}|s_{\alpha\gamma}|^2(f_\alpha{-}f_\gamma)\Big]^2.
     \label{eq:noise_quantum}
\end{align}
\label{app:eq:entropy-noise}
\end{subequations}

\begin{figure}[t]
\includegraphics[width=\columnwidth]{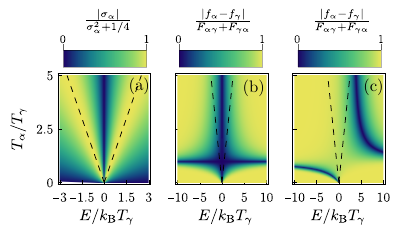}
\caption{\label{fig:ineq} Illustration of the inequalities~\eqref{eq:inequalities} for simple Fermi functions. We set $\mu_\alpha\equiv0$ as the reference energy and plot the ratios of the left and right hand side of Eqs.~\eqref{eq:inequalities} as function of energy and of the ratio of temperatures $T_\alpha/T_\gamma$. 
(a) Plot of $|\sigma_\alpha|/[\sigma_\alpha^2+1/4]$ demonstrating that Eq.~\eqref{eq:log_inequality} is saturated around $E-\mu_\alpha=\pm k_\text{B}T_\alpha/2$, as highlighted by the dashed lines. (b-c) Plots of $|f_\alpha-f_\gamma|/(F_{\alpha\gamma}+F_{\gamma\alpha})$ for (b) $\mu_\gamma=0$ and for (c)  $\mu_\gamma/k_\text{B}T_\gamma=3$. 
The ratio approaches 1 at large energies, when $f_\gamma$ is close to $0$ or $1$ (being at the same time different from $f_\alpha$). 
For comparison, the dashed lines indicate where inequality \eqref{eq:log_inequality} is approached, see panel (a). Consequently, having regions where equality is reached in both Eqs.~\eqref{eq:inequalities} requires a shift of the  chemical potentials between $f_\alpha$ and $f_\gamma$.
{Note that the results in (a) are independent of $T_\gamma$, but we plot the axes in units of $T_\gamma$ to allow one to compare that plot with the plots in (b) and (c).}}
\end{figure}

To establish bounds on the entropy production imposed by these fluctuations, we use the following inequalities that are valid independently of whether the distributions $f_\alpha,f_\gamma$ are thermal or not\footnote{{These inequalities are easily proven as follows. Recalling that $F_{\alpha\gamma}=f_\alpha(1-f_\gamma)$ and $0\leq f_\alpha\leq 1$, one has $0\leq F_{\alpha\gamma}\leq 1$. Therefore, $F_{\alpha\gamma}+F_{\gamma\alpha}\ge |F_{\alpha\gamma}-F_{\gamma\alpha}|$, yielding inequality~\eqref{eq:f_alpha_inequality}. For~\eqref{eq:log_inequality}, one notices that $x^2-|x|+1/4=(2|x|-1)^2/4\ge 0\,\forall x\in\mathbb{R}$, and in particular for $x=\sigma_\alpha$.}}:
\begin{subequations}
\begin{align}
        \left|\sigma_\alpha\right| & \leq \left[\sigma_\alpha\right]^2+\frac14 \label{eq:log_inequality}\,,\\
     |f_\alpha- f_\gamma| & \leq F_{\alpha\gamma} + F_{\gamma\alpha} \ .\label{eq:f_alpha_inequality}
\end{align}
\label{eq:inequalities}
\end{subequations}
To get some intuition about these inequalities, we  illustrate in Fig.~\ref{fig:ineq} under which conditions they turn into equalities, for the simple case where the functions $f_\alpha$ and $f_\beta$ are Fermi functions. 
Using~\eqref{eq:inequalities}, we find that the \textit{classical} component of the noise satisfies
\begin{align}
\label{eq:classical-noise-bound}
     S_{\alpha\alpha,\text{cl}}^{\Sigma} {+} \frac{k_\text{B}^2}{4}S_{\alpha\alpha,\text{cl}}^{}
     &\geq \frac{2k_\text{B}^2}{h}\int dE \left|{\sigma_\alpha}\sum_{\gamma} |s_{\alpha\gamma}|^2(f_\alpha-f_\gamma)\right|\nonumber \\
     &\geq 2k_\text{B}|\Isigma_\alpha|.
\end{align}
This is the first key result of this paper; it relates the entropy production to a \textit{combination} of the classical contributions to entropy and particle fluctuations, which reduces to a combination of the classical contributions to heat and charge fluctuations in the case of thermal contacts.
We are in particular interested in the fluctuations of the nonthermal resource, where the relation~\eqref{eq:classical-noise-bound} provides us with a relevant definition of the classical part of \emph{\rfs}, 
\begin{equation}
    \Scl\equiv S_\text{NN,cl}^\Sigma+\frac{k_\text{B}^2}{4}S_\text{NN,cl}^{}\,,
\end{equation}
that will be used later to characterize the N-demon. We will show in Sec.~\ref{sec:approaching} under which conditions the bound set by the generally valid inequality~\eqref{eq:classical-noise-bound} can be approached in a concrete device realization, even when in addition the strict demon conditions~\eqref{eq:demon-conditions} are imposed.

As a next step, we explore the implications of the constraint~\eqref{eq:classical-noise-bound} on the \rfs of a multi-terminal system for a \textit{useful} device, namely a device associated with negative entropy production in the working substance. If we impose  $\Isigma_\text{ws}<0$, by the second law we have that the entropy production in the resource contact is positive and satisfies $\Isigma_\text{N}\ge -\Isigma_\text{ws}$. Therefore, the inequality~\eqref{eq:classical-noise-bound}, applied to the nonthermal demon terminal, becomes
\begin{equation}
   \Scl= S_\text{NN,cl}^\Sigma+\frac{k_\text{B}^2}{4}S^{}_\text{NN,cl}\ge 2k_\text{B}\Isigma_\text{N}\ge -2k_\text{B}\Isigma_\text{ws}.
    \label{eq:classical_bound_3T}
\end{equation}
as introduced in Eq.~\eqref{eq:result_intro}. This shows that the level of entropy reduction in the working substance, which is an indication of how useful the device is, sets a minimum level of classical fluctuations in the resource terminal. {Equivalently, this means that knowing the amount of classical fluctuations in the resource region, allows one to predict the upper bound of useful entropy reduction that can be achieved in the working substance.}
{In addition, we notice that the first of the above inequalities can be used to put a lower bound on the efficiency of the nonthermal machine. Indeed, one can define a thermodynamically consistent efficiency as $\eta=-\Isigma_\mathrm{ws}/\Isigma_\mathrm{N}$~\cite{Hajiloo_2020}. Then, by using the first inequality in \eqref{eq:classical_bound_3T}, we readily find $\eta\ge -2\kB\Isigma_\mathrm{ws}/\Scl$.}

Intriguingly, Eq.~\eqref{eq:classical_bound_3T} bears some similarities to thermodynamic uncertainty relations (TURs)~\cite{Barato2015Feb,Falasco2020May,Horowitz2020Jan,Vo2022Sep}, while actually containing different information. In its simplest form a TUR states
\begin{equation}
    \frac{\text{Var}(\mathcal{I})}{\Braket{\mathcal{I}}^2}\ge\frac{2}{\dot{\Sigma}}\,,
    \label{eq:TUR}
\end{equation}
where $\mathcal{I}$ is any given current and $\dot{\Sigma}$ is the global entropy production rate. 
Of course, violations of TURs are well-known~\cite{Brandner2018PRL,agarwalla_assessing_2018,saryal_thermodynamic_2019,kalaee_violating_2021,Prech2023Jun}, particularly for quantum systems (as in this work) described by scattering theory~\cite{Potanina2021Apr,Timpanaro2023Mar,Taddei2023} or in multiterminal configurations~\cite{dechant_multidimensional_2018,lopez_optimal_2022}, but that is not the subject here.
Rather, we point out that if we rewrite the first inequality in Eq.~\eqref{eq:classical_bound_3T} as
\begin{equation}
    \frac{S_\text{NN,cl}^\Sigma}{\left(\Isigma_\text{N}\right)^2}\ge\frac{2k_\text{B}}{|\Isigma_\text{N}|}-\frac{k_\text{B}^2S^{}_\text{NN,cl}}{4\left(\Isigma_\text{N}\right)^2}\,,
    \label{eq:inequality_TUR_form}
\end{equation}
then it is very reminiscent of the TUR in Eq.~\eqref{eq:TUR} applied to entropy fluctuations (for which $\Braket{\mathcal{I}}$ in \eqref{eq:TUR} is replaced by the rate of entropy production itself). However,
this similarity is deceptive; our relation~\eqref{eq:inequality_TUR_form} involves the \emph{local} entropy production in a given contact, when the TUR involves global entropy production. Moreover, violations of the TUR~\eqref{eq:TUR} being of quantum origin~\cite{Brandner2018PRL}, they cannot be associated with the additional term in Eq.~\eqref{eq:inequality_TUR_form}, which is an inequality for the classical part of the fluctuations alone. Therefore, despite the appealing similarity between Eqs.~\eqref{eq:TUR} and \eqref{eq:inequality_TUR_form}, we underline that these two relations are different statements.

Until now, we have considered constraints imposed by the classical part of the fluctuations, only. Indeed, the quantum component of the noise is negligible in the tunneling regime, i.e. $|s_{\alpha\gamma}|^2\ll1$, $\alpha\neq\gamma$, but also vanishes in the case where $\alpha$ is a dephasing probe, i.e., when its distribution satisfies $\sum_{\gamma}|s_{\alpha\gamma}|^2(f_\alpha-f_\gamma)=0$ at every energy $E$~\cite{deJong1996Aug}. In general it is however nonzero and, importantly, leads to a reduction of the total noise.

Hence, in order to predict constraints on the full fluctuations associated with the engine's resource, a statement on the \textit{sum} of both the classical and the quantum component is required. In order to do so, we first define $\Tilde{f} \equiv \sum_{\gamma\neq\alpha} |s_{\alpha\gamma}|^2f_\gamma/\sum_{\gamma\neq\alpha} |s_{\alpha\gamma}|^2$ 
and notice that $(f_\alpha - \tilde{f})\in[-1, 1]$. Using this observation (and the unitarity of the scattering matrix), we find the inequalities
\begin{equation}
\begin{split}
    S^\Sigma_{\alpha\alpha, \text{qu}} &\geq -\frac{2k_\text{B}^2}{h}\!\int\!dE[\sigma_\alpha]^2{\left[1-|s_{\alpha\alpha}|^2\right]}^2\left|f_\alpha -\tilde{f}\right|\\
    &\geq -\frac{2k_\text{B}^2}{h}\!\int\!dE[\sigma_\alpha]^2{[1{-}|s_{\alpha\alpha}|^2]}\sum_{\gamma}
    |s_{\alpha\gamma}|^2|f_\alpha-f_\gamma|\ .
\end{split}
\end{equation}
This means that the total entropy fluctuations in contact $\alpha$ satisfy
\begin{equation}
    S_{\alpha\alpha}^\Sigma \geq \frac{2k_\text{B}^2}{h}\int dE\,[\sigma_\alpha]^2|s_{\alpha\alpha}|^2\sum_{\gamma}|s_{\alpha\gamma}|^2|f_\alpha-f_\gamma|.
    \label{eq:bound_full_intermediate}
\end{equation}
Using the inequality in Eq.~\eqref{eq:log_inequality}, we therefore get
\begin{equation}\label{eq:fullbound_inf}
    S_{\alpha\alpha}^\Sigma + \frac{k_\text{B}^2}{4}S_{\alpha\alpha}^{}\geq 2k_\text{B}|\Isigma_\alpha|\inf_{E\in A} |s_{\alpha\alpha}(E)|^2,
\end{equation}
where $A\subseteq\mathbb{R}$ is the support of the integrand function in Eq.~\eqref{eq:bound_full_intermediate}.
This second key result of this section provides a similar inequality to that on the classical noise components in Eq.~\eqref{eq:classical-noise-bound}, but now for the full fluctuations. 
Note that this inequality is less constraining. In particular, it does not provide relevant information when the infimum of $|s_{\alpha\alpha}(E)|^2$ is zero---since the total noise always has to be non-negative---but only when the reflection probabilities are nonzero in the transport window. 
In particular, in the special case where $|s_{\alpha\alpha}(E)|^2=|\bar{s}_{\alpha\alpha}|^2$ is constant and considering $\alpha=\text{N}$, we find
    \begin{equation}
    S_\text{NN}^\Sigma+\frac{k_\text{B}^2}{4}S^{}_\text{NN}\ge 2k_\text{B}|\bar{s}_\text{NN}|^2\Isigma_\text{N}\ge -2k_\text{B}|\bar{s}_\text{NN}|^2\Isigma_\text{ws},
    \label{eq:bound_full}
\end{equation}
setting a minimal constraint on the total noise of the nonequilibrium resource given a certain performance goal in the working substance. Note that in the tunneling regime, we have $|\bar{s}_\text{NN}|^2\approx 1$ and the bound~\eqref{eq:classical_bound_3T} on the classical noise components is recovered, as expected. Furthermore, we see that the quantum noise component lowers the bound, as anticipated. In fact, $|\bar{s}_\text{NN}|^2$ needs to be smaller than 1 to have transport between terminal N and the working substance. 

The bounds~\eqref{eq:classical_bound_3T} and~\eqref{eq:bound_full} become particularly appealing for a three-terminal N-demon, where, thanks to the demon conditions, the entropy reduction $-\Isigma_\text{ws}$ can be written in terms of the output power.\footnote{Extensions to hybrid, multi-functional engines, can be done in line with Ref.~\cite{Manzano2020Dec}.} Let us first consider a system operating as a refrigerator, setting $T_2<T_1$ and $\mu_1=\mu_2=\mu_0$. Then, we have
\begin{equation}
\begin{split}
-\Isigma_\text{ws}&=-\frac{J_1}{T_1}-\frac{J_2}{T_2}=\left(\frac{1}{T_1}-\frac{1}{T_2}\right)(\Ienergy_2-\mu_0 I_2)\\
&=\left(\frac{1}{T_2}-\frac{1}{T_1}\right)J_\text{cool},
\end{split}
\end{equation}
where charge- and energy-current conservation and the demon conditions, Eq.~\eqref{eq:demon-conditions}, have been used. Furthermore,  $J_\text{cool}$ is the cooling power, namely the heat current carried away from the cold contact 2.
If the system runs instead as an engine, we set $T_1=T_2=T_0$ and $\mu_1\neq\mu_2$. Then, using again the demon conditions, we get
\begin{equation}
    -\Isigma_\text{ws}=\frac{\mu_1-\mu_2}{T_0}I_1=\frac{P}{T_0}\,,
\end{equation}
where $P$ is the power produced by the engine (current against a potential bias). 
In summary, we conclude that in a three-terminal N-demon the useful output quantity produced in the working substance (electric power or cooling power) sets the minimum amount of fluctuations (a combination of charge and entropy fluctuations) that must go along with the resource. In the following, we will analyze the impact of the general bounds derived in the Section on concrete N-demon realizations.

\section{Noise bounds for a quantum-Hall N-demon}
\label{sec:minimal-model}

While the bounds~\eqref{eq:classical_bound_3T} and~\eqref{eq:bound_full} are valid for any multi-terminal coherent conductor, we now focus on the situation where the resource terminal is characterized by a given nonthermal distribution as it might arise in nanoscale devices due to the coupling to different environments while thermalization is not effective. We furthermore now request that on average no charge and energy currents flow between resource contact and working substance, namely the strict conditions~\eqref{eq:demon-conditions} are fulfilled. This is of interest since it highlights the nonthermal property of the distribution as an additional resource, distinct from (average) heat flow.
In this section, we analyze a three-terminal configuration, as it is typical for an engine, and choose an implementation relying on a conductor in the quantum Hall regime, depicted in Fig.~\ref{fig:setup_3T}(b), similar to previous analyses~\cite{Sanchez_2019,Hajiloo_2020}.

\subsection{Minimal 3-terminal N-demon configuration}\label{sec:minimal-model_model}
We consider the three-terminal quantum Hall setup shown in Fig.~\ref{fig:setup_3T}(b), where the nonequilibrium distribution $f_\mathrm{N}(E)$ is injected into the working substance through an interface with transmission probability $\tau_\mathrm{int}(E)$. 
The minimal configuration to extract work or to realize cooling in the two-terminal working substance in this chiral conductor is via a scattering region in front of one of these two terminals, with a transmission probability $\tau_\mathrm{ws}(E)$ that has to be energy-dependent~\cite{Hajiloo_2020}.

The demon conditions~\eqref{eq:demon-conditions} for charge current $I_\mathrm{N}=0$ and energy current $\Ienergy_\mathrm{N}=0$ in this setup take the simple form
\begin{equation}
    \left(\begin{array}{c} I_\text{N}\\ \Ienergy_\text{N}\end{array}\right)=\frac{1}{h}\int dE\,\left(\begin{array}{c}1\\ E\end{array}\right)\tau_\text{int}(f_2-f_\text{N})=0.
\label{eq:3T_demoncondition}
\end{equation}
These conditions are minimally affected by the properties of the working substance, since the backflow towards the resource region only depends on $f_2$ and $\tau_\text{int}$. Hence, $T_2$ and $\mu_2$ are the only parameters of the working substance that are relevant, while the concrete realization of the working substance encoded in $\tau_\mathrm{ws}$ is not. The conditions of Eq.~(\ref{eq:3T_demoncondition}) can hence be fulfilled by adjusting the nonthermal distribution $f_\mathrm{N}$, independently of $\tau_\mathrm{ws}$.

In the working substance of this setup, the average energy current reads
\begin{subequations}
\begin{align}
    \Ienergy_1&\!=\frac{1}{h}\int dE\,E\,\tau_\text{ws}[(f_2-f_1)+\tau_\text{int}(f_\text{N}-f_2)],\label{eq:currents_3T_1}\\
    \Ienergy_2&\!=\frac{1}{h}\int dE\,E\,[\tau_\text{ws}(f_1-f_2)+\tau_\text{int}(1-\tau_\text{ws})(f_\text{N}-f_2)].\label{eq:currents_3T_2}
\end{align}
\label{eq:currents_3T}
\end{subequations}
and equivalently for the particle currents ($I_1$ and $I_2$), where the energy factor in the integral is replaced by a 1. These currents can lead to refrigerator and engine operation of the setup under demon conditions, provided that the entropy production
\begin{equation}
    \Isigma_\text{N}=\frac{k_\text{B}}{h}\int dE\,\sigma_\mathrm{N}\tau_\text{int}(f_2-f_\text{N})
    \label{eq:entropy-demon_3T}
\end{equation}
in the resource terminal is positive.
However, as we saw in Sec.~\ref{sec:generalConstraints}, there is a minimum amount of noise in the zero-average flow from the N-terminal required for the system to produce a useful output. 
For the setup of Fig.~\ref{fig:setup_3T}(b), we here show that the full fluctuations of both $I_\text{N}$ and $\Ienergy_\text{N}$ have to be non-zero even separately, meaning that these currents (both with zero average) are necessarily \textit{both} noisy. 
We prove this by showing that requiring no fluctuations prevents the generation of any useful output. A straightforward calculation of the energy noise gives
\begin{equation}
    S_\text{NN}^\mathrm{E}=\frac{2}{h}\int \!dE\,E^{2}\tau_\text{int}[F_\text{NN}+F_{22}+(1-\tau_\text{int})(f_2-f_\text{N})^2]\,,
\label{eq:noise_demon_3T}
\end{equation}
and equivalently for the particle current noise $S_\mathrm{NN}$ by dropping the factor $E^2$ in the integrand. An obvious possibility to nullify the noise is to set $\tau_\text{int}=0$, but this means that the working substance and the resource region are completely decoupled\footnote{This is in contrast to Coulomb-coupled systems, where energy and information exchange can take place in the absence of particle exchange~\cite{Sanchez2011Feb,Strasberg2013Jan,koski_experimental_2014,Whitney2015,Sanchez2019Oct}.}. As a result, the resource region cannot have any effect on the working substance, where therefore all currents flow according to the direction imposed by the voltage and temperature biases between contacts 1 and 2. 
For $\tau_\mathrm{int}\neq 0$, all three terms in the square bracket in Eq.~(\ref{eq:noise_demon_3T}) have to vanish separately and for every energy (since they can never be negative).
Examining the term $F_{22}=f_2(1-f_2)$ in the noise and taking into account that $f_2$ is a thermal distribution, we must require $T_2=0$ for $F_{22}$ to vanish. We now examine what this implies for the demon conditions.
Taking $\mu_2$ as the reference energy, one {sees that $T_2=0\implies E\,\tau_\mathrm{int}(f_2-f_\mathrm{N})\le 0\implies \Ienergy_\text{N}\le 0$.} Hence, the only way to satisfy the demon condition, Eq.~(\ref{eq:3T_demoncondition}), is by imposing $\tau_\text{int}(f_2-f_\text{N})=0$ for all energies. 
As a result, the currents in the working substance reduce to $hI_2^{}=\int dE\,\tau_\text{ws}(f_1-f_2)=-hI_1^{}$ and $h\Ienergy_2=\int dE\,E\tau_\text{ws}(f_1-f_2)=-h\Ienergy_1$, so that the flows are only determined by the temperature and voltage biases of contacts 1 and 2. 
Thus, one concludes that the presence of charge \textit{and} energy fluctuations of the incoming (zero-average) flux from the N-terminal is essential for the functioning of the N-demon. Importantly, we have shown this for the \textit{total} fluctuations---namely the sum of classical and quantum contributions---which in the case where $\tau_\mathrm{int}=1$ would not be excluded to vanish if one considers  Eq.~\eqref{eq:fullbound_inf}.

\subsection{Conditions to approach the fluctuations bound}\label{sec:approaching}
We have seen in Fig.~\ref{fig:ineq} how the inequalities~\eqref{eq:inequalities} can be saturated by shifting Fermi functions with respect to each other.  
In this section, we explore to which extent it is possible to approach the fluctuations bound~\eqref{eq:classical_bound_3T} resulting from these inequalities for the 3-terminal system introduced in Fig.~\ref{fig:setup_3T}(b). Hence, we exploit a nonthermal resource that can strongly differ from a Fermi function, and aim at carrying out a useful task under demon conditions. 

We start by looking at the conditions for which one obtains an equality in relations~\eqref{eq:inequalities}, in analogy to what was obtained in Fig.~\ref{fig:ineq}. The inequality~\eqref{eq:log_inequality} is saturated for a (piecewise) constant distribution $f_\alpha$, that takes either of the two optimal values
\begin{equation}
    f_\pm\equiv \frac{1}{1+e^{\mp 1/2}}.
    \label{eq:optimal_values}
\end{equation}
Instead, reaching equality in relation~\eqref{eq:f_alpha_inequality} requires $f_\gamma\in\{0,1\}$. Specifying to the setup in Fig.~\ref{fig:setup_3T}(b), we therefore see that saturating the first part of the inequality~\eqref{eq:classical-noise-bound} requires $f_2\in\{0,1\}\implies T_2=0$ and $f_\text{N}\in\{f_+,f_-\}$.
However, the first of these constraints cannot be satisfied by imposing, at the same time, the demon conditions~\eqref{eq:demon-conditions} and that the N-demon produces a useful output, as shown in Sec.~\ref{sec:minimal-model_model}. 
Therefore, we come to the important conclusion that requiring a useful N-demon does not allow one to \emph{reach} any of the two bounds~\eqref{eq:classical_bound_3T} and~\eqref{eq:bound_full}. 
We analyze in the following which configurations allow one to \textit{approach} the bound and to what extent.

\begin{figure}[t]
\includegraphics[width=0.75\columnwidth]{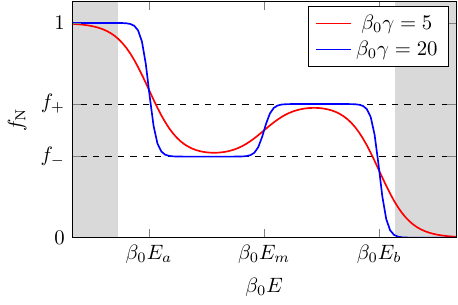}
\caption{\label{fig:sketch_fN_optimal}
Example of the nonequilibrium distribution given in Eq.~\eqref{eq:noneq_bound_approach}, with $\beta_0 E_a=-2$, $\beta_0 E_b=1$, $E_m=(E_a+E_b)/2$, and two different values of the sharpness $\gamma$. The dashed horizontal lines indicate the values $f_+\approx 0.62$ and $f_-=1-f_+\approx 0.38$, cf.~Eq.~\eqref{eq:optimal_values}. The non-shaded region corresponds to the selected energy interval with $\Delta = 5$.
Energies are taken with respect to a common chemical potential $\mu=0$ and a reference temperature $T_0.$}
\end{figure}

We now illustrate with an example that it is indeed possible that the level of fluctuations of a working N-demon gets reasonably close to the bound.
We therefore first lift the constraint $T_2=0$ in order to allow for a finite power output under demon conditions. Next, it is natural to require the nonthermal distribution $f_\text{N}$ to take values that are close to the optimal ones, given in Eq.~\eqref{eq:optimal_values}. While there are many ways to fulfill this condition, we here focus on the simple case where $f_\text{N}$ is a function of the form
\begin{equation}
\begin{split}
    f_\text{N}(E)&=\vartheta_\gamma(E_a-E)+f_-\vartheta_\gamma(E-E_a)\\
    &+(f_+-f_-)\vartheta_{\gamma}(E-E_m)-f_+\vartheta_\gamma(E-E_b),
\end{split}
\label{eq:noneq_bound_approach}
\end{equation}
with $E_a<E_m<E_b$ and
\begin{equation}
    \vartheta_\gamma(E)=\frac{1}{1+\exp(-\gamma E)},
\end{equation}
see Fig.~\ref{fig:sketch_fN_optimal} for an illustration. This function~\eqref{eq:noneq_bound_approach} transitions between the values $1, f_-, f_+,0$ with increasing energy, and the parameter $\gamma$ governs the sharpness of the transitions.
Note that having $f_\text{N}$ approach 0 or 1 is not beneficial for reaching the equality in Eq.~\eqref{eq:bound_full}, but is physically relevant. Indeed, even in conductors with nonthermal distributions, we expect states at very high energies to be rarely filled and states at very low energies to be rarely empty.
Moreover, we choose the interface transparency, $\tau_\mathrm{int}(E)$, in such a way that all the energy integrals are limited to the interval $(E_a-|E_a|/\Delta, E_b+|E_b|/\Delta)$, thus keeping energy regions where the nonequilibrium distribution can deviate significantly from the optimal values $f_\pm$ (see non-shaded region in Fig.~\ref{fig:sketch_fN_optimal} for an example). Having fixed the integration interval, one can calculate all the transport quantities. In particular, the demon conditions $I_\text{N}=\Ienergy_\mathrm{N}=0$ will be used to fix the values of $E_a$ and $E_b$.

Focusing on the classical part of the \rfs, for which inequality~\eqref{eq:classical_bound_3T} holds, we now study the behavior of the ratio\footnote{{The inverse of this fraction, $1/\mathcal{R}^{3\mathrm{T}}_{\text{cl}}$, can be seen as a type of performance quantifier setting the maximum amount of work that can be done, see also Eq.~(\ref{eq:classical_bound_3T}), given a certain level of fluctuations.}}
\begin{equation}
\mathcal{R}^{3\mathrm{T}}_{\text{cl}}=\frac{S_\mathrm{NN,cl}^{\Sigma} {+} \frac{k_\text{B}^2}{4}S_\mathrm{NN,cl}^{}}{2k_\text{B}|\Isigma_\mathrm{N}|}\ge 1,
\label{eq:bound_ratio}
\end{equation} 
as a function of the transition sharpness $\gamma$ and the parameter $\Delta$ governing how much the allowed energy interval exceeds the boundaries $E_a$ and $E_b$. 

\begin{figure}[t]
\includegraphics[width=0.85\columnwidth]{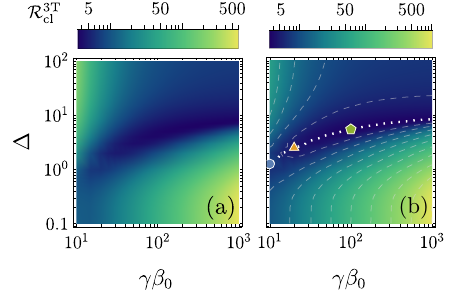}
\caption{\label{fig:bound_ratio_1}
Ratio $\mathcal{R}^{3\mathrm{T}}_{\text{cl}}$ defined in Eq.~\eqref{eq:bound_ratio}, calculated for the nonequilibrium distribution~\eqref{eq:noneq_bound_approach} with $E_m=(E_a+E_b)/2$ and $E_a$, $E_b$ fixed by the demon conditions. The ratio is plotted as a function of the parameters $\gamma$ and $\Delta$. The other parameters are chosen having in mind a refrigerator configuration. Therefore, we set the reference temperature $T_1=T_0$, a common electrochemical potential $\mu_1=\mu_2\equiv 0$, and (a) $T_2=0.3T_0$, (b) $T_2=0.5T_0$. The markers in (b) indicate the points we use to calculate the cooling power of an N-demon using such a nonequilibrium distribution, see Fig.~\ref{fig:cooling_close_bound}.
These points are chosen along the line (white dashed) where $\mathcal{R}^{3\mathrm{T}}_{\text{cl}}$ takes the minimal values as a function of $\gamma$ and $\Delta$.}
\end{figure}

The result is shown in Fig.~\ref{fig:bound_ratio_1}. The dark blue region represents the parameter space where the ratio $\mathcal{R}^{3\mathrm{T}}_{\text{cl}}$ is relatively low (the global minimum in the displayed plot is around 3.5).
The behavior in Fig.~\ref{fig:bound_ratio_1} can be roughly understood as follows. 
Inequality~\eqref{eq:log_inequality} tells us that the bound can be approached when $f_\text{N}$ is close to the optimal values $f_\pm$. By contrast, if $f_\text{N}$ approaches the values 0 or 1, the inequality becomes very inaccurate. Therefore, to achieve a small $\mathcal{R}^{3\mathrm{T}}_{\text{cl}}$ the contributions where $f_\text{N}\approx 0, 1$ need to be suppressed while maintaining the ones where $f_\text{N}\approx f_\pm$. 
As depicted in Fig.~\ref{fig:bound_ratio_1}, this can be achieved in two ways. For large $\Delta$ and large $\gamma$ the regions where $f_\text{N}$ is not close to $f_\pm$ are filtered out: a large value of $\gamma$ makes $f_\text{N}$ transition abruptly between the values 0, $f_+$, $f_-$, 1 (see the blue line in Fig.~\ref{fig:sketch_fN_optimal}), and a large value of $\Delta$ selects the energy region where $f_\text{N}=f_+,f_-$ as integration interval. This explains the dark blue region in the upper right corners of the plots in Fig.~\ref{fig:bound_ratio_1}.
But one can also achieve a small $\mathcal{R}^{3\mathrm{T}}_{\text{cl}}$ value by decreasing the parameters $\Delta$ and $\gamma$, visible on the diagonals of both panels of Fig.~\ref{fig:bound_ratio_1}. 
Decreasing $\gamma$, the steps in $f_\text{N}$ become smoother, see the red line in Fig.~\ref{fig:sketch_fN_optimal}, meaning that $f_\mathrm{N}$ deviates only slightly from the optimal values $f_\pm$ in extended energy intervals. Indeed, Fig.~\ref{fig:ineq}(a) shows that for smooth distribution functions (arising from large temperatures in Fig.~\ref{fig:ineq}(a)) the region where the ratio $|\sigma_\alpha|/(\sigma_\alpha^2+1/4)$ is close to 1 is significantly increased. As a consequence, it is possible to extend the integration interval beyond $[E_a,E_b]$ by decreasing $\Delta$ while staying in a range for which Eq.~\eqref{eq:log_inequality} is sufficiently accurate. Note that detailed features of Fig.~\ref{fig:bound_ratio_1} furthermore heavily rely on the interconnection between the different parameters through the demon conditions, and can hence not straightforwardly be explained by the simple reasoning above. 

\begin{figure}[t]
\includegraphics[width=0.8\columnwidth]{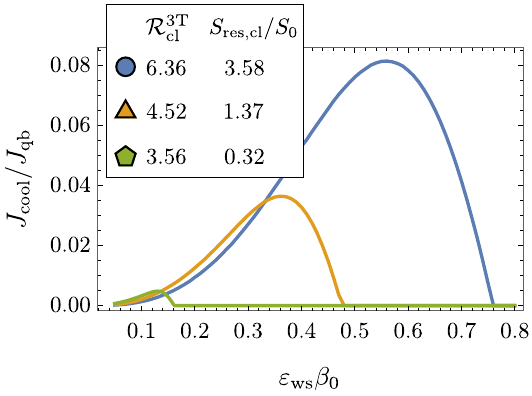}
\caption{\label{fig:cooling_close_bound}
Cooling power $J_\text{cool}$ when the three terminal N-demon is operated as a refrigerator, with $T_1=T_0$, $T_2=0.5T_0$, and $\mu_1=\mu_2=0$. The working substance exploits the incoming nonequilibrium distribution~\eqref{eq:noneq_bound_approach} by using a filter transmission probability $\tau_\text{ws}$ of the form~\eqref{eq:filter_ws}. The three curves correspond to different working points, indicated in Fig.~\ref{fig:bound_ratio_1}(b). The associated values of the ratio $\mathcal{R}^{3\mathrm{T}}_{\text{cl}}$ and the classical resource fluctuations $S_\text{res,cl}$ (in units of $S_0=2k_\text{B}^3T_0/h$) are shown in the legend.}
\end{figure}
Overall, our analysis shows that it is possible to conceive nonthermal distributions allowing one to approach the bounds derived in the previous section, even when satisfying the demon conditions.
It remains to be shown that the N-demon can produce a useful output in these regimes. We illustrate this for the example of the demon acting as a refrigerator cooling contact 2, which is set at temperature $T_2<T_1$. The working substance must be equipped with an energy-dependent transmission probability $\tau_{\text{ws}}(E)$ to be able to exploit the incoming nonthermal resource. For this purpose, we choose an energy filter in the form of a boxcar function:
\begin{equation}
    \tau_\text{ws}(E)=
    \begin{cases}
        1 & -\varepsilon_\text{ws}\le E\le\varepsilon_\text{ws}\,,\\
        0 & \text{otherwise}\,.
    \end{cases}
    \label{eq:filter_ws}
\end{equation}
This choice is motivated by the high relevance of this type of transmission in quantum thermoelectrics~\cite{Whitney2014PRL,Whitney2015}, also concerning noise~\cite{Kheradsoud2019Aug,Timpanaro2023Mar,Balduque2023Jul}. 
The results for the cooling power $J_\text{cool}=-J_2$ obtained in this situation are shown in Fig.~\ref{fig:cooling_close_bound} as a function of the filter half-width $\varepsilon_\text{ws}$. The cooling power is normalized with Pendry's bound~\cite{Pendry1983Jul}
\begin{equation}
    J_\text{qb}=\frac{\pi^2k_\text{B}^2}{6h}T_\text{cold}^2,
    \label{eq:Pendry}
\end{equation}
which sets the maximum amount of heat that can be carried away from a contact with a given temperature, here $T_\mathrm{cold}$. Quite remarkably, it is possible to have a finite cooling power, approaching almost 10\% of Pendry's bound. Moreover, Fig.~\ref{fig:cooling_close_bound} also shows that the curve displaying the largest cooling power corresponds to the largest value of
    $\Scl=S_\text{NN,cl}^\Sigma+\frac{k_\text{B}^2}{4}S_\text{NN,cl}^{}$.
This means that for the example considered here, the classical \rfs grow faster than the entropy production of the N-terminal and the entropy reduction in the working substance---more than imposed by the inequality~\eqref{eq:classical_bound_3T}.

\section{Noise bounds for an engineered multi-terminal N-demon}
\label{sec:composite}

\begin{figure}[t]
\includegraphics[width=\columnwidth]{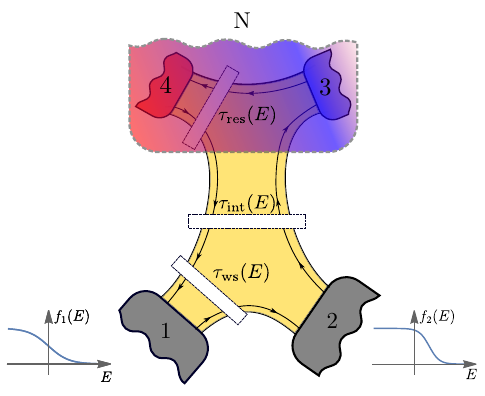}
\caption{\label{fig:setup_4T}
A four-terminal configuration for the N-demon, where the nonthermal resource is implemented by a coherent conductor mixing the occupation probabilities of two equilibrium contacts, 3 and 4, thanks to an energy-dependent transmission function $\tau_\text{res}(E)$ associated with a scatterer in the resource region.}
\end{figure}
Until now, we have considered a device where the resource is a \emph{single} terminal providing a given nonthermal distribution $f_\text{N}$ and we have {purposefully} not introduced a specific mechanism creating this distribution. 
Nonthermal distributions are ubiquitous in nanoscale devices and various examples have been analyzed how to engineer them, such as with squeezed states~\cite{Rossnagel2014Jan,Correa2014Feb,Abah2014May,Manzano2016May,Agarwalla2017Sep,Manzano2018Oct}, quantum correlations~\cite{Francica2017Mar}, or driven contacts~\cite{Ryu2022May}. 

{To give a concrete, but simple example, representative of a system in contact with different environments, yet realizable and tunable in experiment, we analyze in} the following {the setup shown in} Fig.~\ref{fig:setup_4T}. {Here,} the resource consists of a set of two thermal contacts, which can have different temperatures and electrochemical potentials, connected via a coherent conductor characterized by an energy-dependent scatterer, with transmission probability $\tau_\text{res}(E)$. 
This combination of different thermal contacts is an extremely simple way to engineer an effective nonthermal distribution, namely
\begin{equation}
    f_\text{N}(E)=\tau_\text{res}(E)f_4(E)+[1-\tau_\text{res}(E)]f_3(E)\,.
\label{eq:fneq_map}
\end{equation}
Such a configuration, as previously studied in Refs.~\cite{Sanchez_2019,Hajiloo_2020}, is expected to be experimentally realizable while being highly tunable. 
Indeed, nanofabrication techniques in two-dimensional electron gases allow for the realization of complex energy-filtering setups in the quantum Hall regime~\cite{Altimiras2010Jan}. In addition, the parameters of the transmission function $\tau_\text{res}(E)$ can usually be controlled by means of gate voltages, as in the case of a quantum point contact~\cite{buttiker_quantized_1990} or resonant quantum-dot-like structures~\cite{Altimiras2010Jan,Gasparinetti2011May,Gasparinetti2012Jun}. The resource region can hence be engineered such that it constitutes the ``demon'' part of the setup, providing a nonthermal resource at vanishing average charge and energy currents out of the shaded region in Fig.~\ref{fig:setup_4T} and into the working substance. Note that distributions analogous to Eq.~\eqref{eq:fneq_map} can also arise in conductors coupled to different baths and subject to different equilibration/thermalization processes, as demonstrated in Ref.~\cite{Tesser2023Apr} for a hot-carrier solar cell.

\subsection{General noise bounds for the multi-terminal resource}
\label{sec:mixing}

We now analyze how the entropy production in the resource (and hence the entropy reduction in the working substance) is bounded by the fluctuations occurring in the resource terminals. We therefore exploit the previously presented bound~\eqref{eq:classical-noise-bound}, but now consider the \emph{sum of the contributions} from the two thermal contacts that together constitute the demon part of the device. Indeed, using Eq.~\eqref{eq:classical-noise-bound}, we get
\begin{equation}
    S_{33,\text{cl}}^\Sigma+S_{44,\text{cl}}^\Sigma+\frac{k_\text{B}^2}{4}[S_{33,\text{cl}}^{}+S_{44,\text{cl}}^{}]\ge 2k_\text{B}(|\Isigma_3|+|\Isigma_4|).
    \label{eq:sum_demon_fluctuations}
\end{equation}
Note that in this inequality, the sum of the moduli of the entropy productions in the resource terminals appears. If---as desired---the two resource terminals indeed \emph{produce} entropy, this sum equals the total entropy production in the resource. Otherwise, we have $ |\Isigma_3|+|\Isigma_4|\geq |\Isigma_3+\Isigma_4|\geq |\Isigma_1+\Isigma_2|$, the latter being the entropy reduction in the working substance. As a side remark, note that the total entropy production of a nonthermal distribution arising from a coherent mixing as presented here is always larger than the entropy production $\Isigma_\text{N}$ of a single terminal with an analogous distribution function $f_\mathrm{N}$, see Ref.~\cite{Tesser2023Apr}. 

Recalling that terminals 3 and 4 are in a thermal state, the entropy fluctuations reduce to the heat noise $S_{\alpha\alpha}^J$, such that
\begin{eqnarray}\label{eq:classical_bound_4T}
 \Scl^\mathrm{4T} &\equiv&   \frac{S_{33,\text{cl}}^J}{T_3^2}+\frac{S_{44,\text{cl}}^J}{T_4^2}+\frac{k_\text{B}^2}{4}\left(S_{33,\text{cl}}^{}+S_{44,\text{cl}}^{}\right)\nonumber\\
 &\ge& -2k_\text{B}\Isigma_\text{ws}.
\end{eqnarray}
This relation provides us with a relevant definition of \emph{classical \rfs} of the 4-terminal setup, $\Scl^\mathrm{4T}$, that will be used for the device characterization.
This inequality is extended to the full noise, including quantum contributions, by exploiting Eq.~\eqref{eq:bound_full}, yielding
\begin{eqnarray}\label{eq:full_bound_4T}
\Stot^\mathrm{4T} & \equiv & \left[\frac{S_{33}^J}{T_3^2}+\frac{k_\text{B}^2}{4}S_{33}^{}\right]+\left[\frac{S_{44}^J}{T_4^2}+\frac{k_\text{B}^2}{4}S_{44}^{}\right]\nonumber\\
&\ge& -2\inf\{|s_{33}(E)|^2, |s_{44}(E)|^2\}k_\text{B}\Isigma_\text{ws}.
\end{eqnarray}
In the following Section~\ref{sec:analysis}, we analyze how the entropy production and its fluctuations behave in concrete realizations of engines and refrigerators realized in quantum-Hall conductors with a set of two terminals, together acting as a demonic nonthermal resource (without average charge and heat currents into the working substance).

\subsection{Analysis of specific device realizations}\label{sec:analysis}

\begin{figure}[t]
\includegraphics[width=\columnwidth]{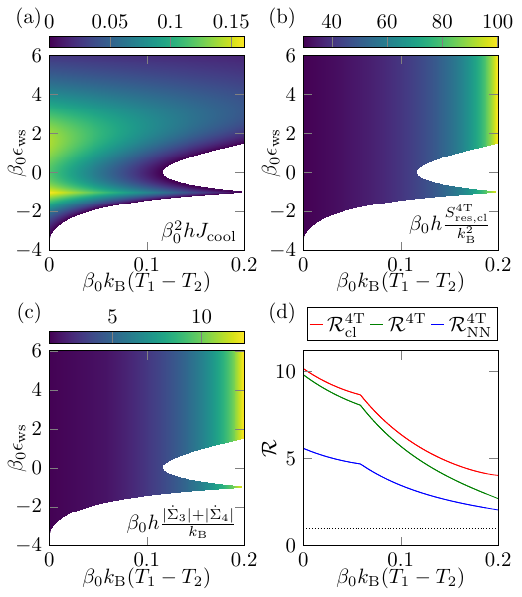}
\caption{\label{fig:refrigerator_config1}
Results for a refrigerator configuration where both $\tau_\text{ws}$ and $\tau_\text{res}$ are sharp step functions, see Eq.~(\ref{eq:QPC}). (a) Cooling power, (b) classical \rfs~$S_\text{res,cl}$, and (c) entropy production $|\Isigma_3|+|\Isigma_4|$ as a function of $\Delta T=T_1-T_2$ and of the position $\epsilon_\text{ws}$ of the step function onset in the working substance.
(d) Ratio between the classical \rfs, classical and quantum \rfs, and fluctuations of the effective N-terminal and the entropy production $2k_\text{B}(|\Isigma_3|+|\Isigma_4|)$. The plotted ratios are independent of  $\epsilon_\text{ws}$. In all panels, we set $\mu_1=\mu_2=0$, $T_1=T_0$, $T_3=T_4=1.2T_0$, and $\beta_0\epsilon_\text{res}=-1$. The white regions in the density plots are those where no cooling is possible under demon conditions. 
}
\end{figure}

\begin{figure}[t]
\includegraphics[width=\columnwidth]{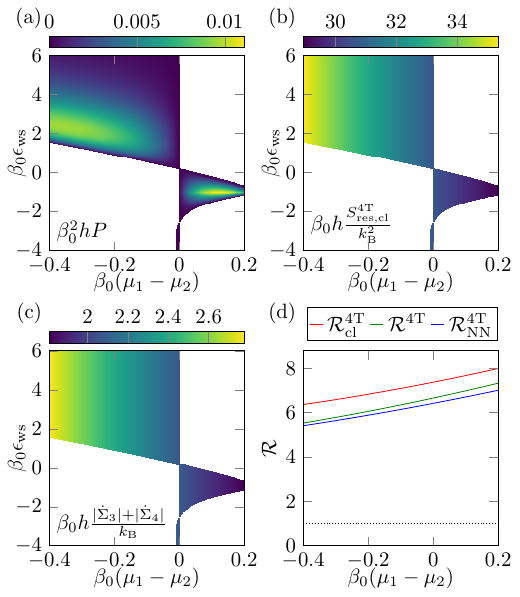}
\caption{\label{fig:engine_config1}
Results for an engine configuration producing electrical power where both $\tau_\text{ws}$ and $\tau_\text{res}$ are sharp step functions, see Eq~(\ref{eq:QPC}). (a) Power, (b) classical \rfs $S_\text{in,cl}$, and (c) entropy production $|\Isigma_3|+|\Isigma_4|$ as a function of $\Delta T=T_1-T_2$ and of the position $\epsilon_\text{ws}$ of the step function onset in the working substance. (d) Ratio between the classical \rfs, classical and quantum \rfs, and fluctuations of the effective N-terminal and the entropy production $2k_\text{B}(|\Isigma_3|+|\Isigma_4|)$. The plotted ratios are independent of  $\epsilon_\text{ws}$. In all panels, we set $T_1=T_2=T_0$, $T_3=T_4=0.7T_0$, and $\beta_0\epsilon_\text{res}=-1$\,. The white regions in the density plots are those where no power is produced  under demon conditions.}
\end{figure}

Here, we characterize the performance of the device in Fig.~\ref{fig:setup_4T} with respect to the bounds we have previously derived. We consider both a refrigerator and an engine configuration. In all cases, we choose to fulfill the demon conditions \eqref{eq:demon-conditions} by adjusting the chemical potentials $\mu_3$ and $\mu_4$ of the resource region. The free parameters of the resource region are thus the temperatures $T_3$ and $T_4$, as well as the transmission function $\tau_\text{res}$. We furthermore consider here $\tau_\text{int}=1$, whereas $\tau_\text{ws}$ and $\tau_\text{res}$ are both chosen to be sharp step functions, which can be obtained as a limiting case of a QPC transmission \cite{vanWees1988Feb,buttiker_quantized_1990,vanHouten1992Mar,Kheradsoud2019Aug}
\begin{equation}
    \tau_\text{QPC}(E)=\frac{1}{1+\exp[-2\pi\gamma_\mathrm{QPC}(E-\epsilon)]}\,,\label{eq:QPC}
\end{equation}
with $\gamma_\mathrm{QPC}$ larger than all relevant energy scales (temperatures, biases) in the system.
We address two types of engine operation: In Fig.~\ref{fig:refrigerator_config1}, we analyze the system working as a refrigerator, where the working substance is characterized by $T_2<T_1$ and $\mu_1=\mu_2=0$. The nonthermal resource is exploited to obtain a finite cooling power in the resource, namely a heat current flowing \emph{out of} the cold contact 2 and into the hotter contact 1.
In Fig.~\ref{fig:engine_config1}, we analyze the system working as an engine producing electrical power, where the working substance is characterized by $T_1=T_2$. Power is produced by driving a particle current \emph{against} the potential difference $\mu_1-\mu_2$.

In these figures, we plot in panels (a) the desired output, namely the cooling power and the produced electrical power. We also plot the classical \rfs, i.e., the combination of classical entropy and particle current fluctuations as defined by Eq.~\eqref{eq:classical_bound_4T}, in panels (b). Furthermore, we show the sum of the moduli of the entropy production in the two resource terminals (ideally corresponding to the total entropy production in the resource), namely $|\Isigma_3| + |\Isigma_4|$, in panels (c). Finally, in panels (d) of Figs.~\ref{fig:refrigerator_config1} and \ref{fig:engine_config1}, we plot the ratios between \rfs and entropy production. We compare the following ratios
\begin{subequations}\label{eq:ratios}
    \begin{align}
        \mathcal{R}_\text{cl}^\text{4T} = \frac{1}{2k_\text{B}}\frac{\Scl^\text{4T}}{|\Isigma_3| +|\Isigma_4|},\label{eq:ratio_classical_input}\\
        \mathcal{R}^\text{4T} = \frac{1}{2k_\text{B}}\frac{\Stot^\text{4T}}{|\Isigma_3| +|\Isigma_4|},\label{eq:ratio_total_input}\\
        \mathcal{R}_\text{NN}^\text{4T} = \frac{1}{2k_\text{B}}\frac{S_\text{res,NN}^\text{4T}}{|\Isigma_3| +|\Isigma_4|}.\label{eq:ratio_N_input}
    \end{align}
\end{subequations}
For the first ratio,  we know  from Eq.~\eqref{eq:sum_demon_fluctuations} that $\mathcal{R}_\text{cl}^\text{4T}\geq 1$.  
By contrast, the second ratio involving the full \rfs, can become smaller than 1 due to the quantum noise contribution, which can be negative. 
Indeed, in the considered setup, where $|s_{33}(E)|^2,|s_{44}(E)|^2$ can become zero, the inequality~\eqref{eq:full_bound_4T} is uninformative, only requiring the trivial fact that the total noise is positive, i.e., $\mathcal{R}^\text{4T}\geq 0$.
We compare these two ratios to a third one, involving the full fluctuations of the zero-average currents flowing from the two-terminal resource into the working substance, $\hat{I}_\text{N}^{} = \hat{I}_3^{} + \hat{I}_4^{}$ and $\hat{I}_\text{N}^\mathrm{E} = \hat{I}_3^\mathrm{E} + \hat{I}_4^\mathrm{E}$.  
The fluctuations appearing in Eq.~\eqref{eq:ratio_N_input}, given by the combination $S_\text{res,NN}^\text{4T} \equiv S_\text{NN}^{\text{4T},\Sigma} +\kB^2 S_\text{NN}^{\text{4T}}/4$, hence also contain the cross-correlation terms between currents $\hat{I}_3^{X}$ and $\hat{I}_4^{X}$, namely
$S_\text{NN}^{\text{4T},X} \equiv S_{33}^{X} + S_{44}^{X} + S_{43}^{X} + S_{34}^{X}$. This is one further possibility to quantify the fluctuations related to the nonthermal resource. Note however, that while being an experimentally relevant quantity, it does not unambiguously quantify the fluctuations of a \textit{generic} nonthermal resource described by the distribution function $f_\mathrm{N}$, as it might seem. 

Indeed, while the average particle and energy currents $I_\text{N}^{}$ and $I_\text{N}^\mathrm{E}$ flowing from the two-terminal resource considered in this section correspond to the average currents flowing from a single nonthermal resource with an equivalent distribution $f_\text{N}$ in Eq.~\eqref{eq:fneq_map}, see  Eq.~\eqref{eq:3T_demoncondition}, the noise of the 4-terminal system 
\begin{equation}
\begin{split}
S_\text{NN}^\text{4T,E}=\frac{2}{h}\int dE\,&E^{2}\tau_\text{int}[F_{22}+(1-\tau_\text{res})F_{33}+\tau_\text{res}F_{44}\\
&+(1-\tau_\text{int})(1-\tau_\text{res})(f_2-f_3)^2\\
&+\tau_\text{res}(1-\tau_\text{int})(f_2-f_4)^2\\
&+\tau_\text{int}\tau_\text{res}(1-\tau_\text{res})(f_3-f_4)^2],
\end{split}
\label{eq:noise_demon_4T}
\end{equation}
is generally different from the one of the 3-terminal system $S_\text{NN}^\mathrm{E}$ in Eq.~\eqref{eq:noise_demon_3T} (and equivalently for the particle current noise). The reason for this is that the noise is nonlinear. Hence, the fluctuations of the effective N-terminal $S_\text{res,NN}^{\text{4T}}$ do not generally satisfy the bound in Eq.~\eqref{eq:full_bound_4T}, as demonstrated in Appendix~\ref{app:current-fluctuations}. 

In the following, we discuss the described quantities for the realization of the N-demon for refrigeration and power production.

\subsubsection{N-demon for cooling}

We start by presenting results for the N-demon operating as a refrigerator. Here, we exploit the nonthermal resource to cool contact 2 in the working substance. 
The results are presented in Fig.~\ref{fig:refrigerator_config1}.
We first notice that it is possible to obtain a finite cooling power even when the temperatures of the resource contacts are higher than the temperatures in the working substance, here for $T_1=T_0$, $0.8T_0<T_2<T_0$, and $T_3=T_4=1.2T_0$. 
In contrast to absorption refrigerators, where refrigeration is driven by a hot contact~\cite{Segal2018May,Mitchison2019Apr,manikandan_autonomous_2020}, this however does not lead to any energy exchange on average between resource and working substance in the situation studied here. 
In Fig.~\ref{fig:refrigerator_config1}(a), we find two maxima in the cooling power $J_\mathrm{cool}$ as function of $\epsilon_\mathrm{ws}$. 
One of them is found at
$\epsilon_\mathrm{ws}=\epsilon_\mathrm{res}=-k_\mathrm{B}T_0$, where the filtering in the resource region and in the working substance takes place at the same energy. The other one is situated around $\epsilon_\mathrm{ws}\approx2 k_\mathrm{B}T_0$; which of the two features is the global maximum depends on the temperature difference in the working substance and changes approximately at the place where the entropy production in contact 3 is found to change sign, see also discussion of panel (d). 

The classical \rfs, $\Scl^\text{4T}$, in panel (b), and the entropy production, $|\Isigma_3| +|\Isigma_4|$, in panel (c), show similar features: they are both increasing with the temperature bias $T_1-T_2$ and are independent of the working substance transmission, here parametrized through $\epsilon_\text{ws}$. 
This independence of the working-substance implementation is an important difference compared to standard Maxwell demons, where the entropy production of the demon is typically directly related to the entropy reduction in the working substance, see also Ref.~\cite{Sanchez_2019}.
We compare the fluctuations and the entropy production in the resource region through the ratios in Eq.~\eqref{eq:ratios} in panel (d). 
As $\mathcal{R}_\text{cl}^\text{4T}$, $\mathcal{R}^\text{4T}$, $\mathcal{R}_\text{NN}^\text{4T}$ each characterize the resource region, they are independent of $\epsilon_\text{ws}$ in the considered setup, see Fig.~\ref{fig:setup_4T}. 
First, we notice that the ratio $\mathcal{R}_\text{cl}^\text{4T}\geq1$ as required by the inequality \eqref{eq:sum_demon_fluctuations}.
Furthermore, panel (d) shows that $\mathcal{R}_\text{cl}^\text{4T}\geq \mathcal{R}^\text{4T}$, which is consistent with the quantum noise contribution being negative.
Moreover, we also observe that $\mathcal{R}^\text{4T}\geq \mathcal{R}_\text{NN}^\text{4T}$. While this may in principle be not true (due to the possibility that heat current cross-correlations are positive even for fermionic systems), it turns out to be the case in the analyzed setup.
Interestingly, each ratio lies above 1, indicated by the dotted line, which is neither required for the total  \rfs, $\Stot^{4\mathrm{T}}$, by the inequalities~\eqref{eq:full_bound_4T}, nor for the effective N-terminal fluctuations, $S_\mathrm{res,NN}^{4\mathrm{T}}$, by relation~\eqref{eq:sum_demon_fluctuations} together with Eq.~\eqref{eq:engineered_bounds}. 
Nonetheless, this feature can be attributed to the fact that, while the  considered setup has a resource fluctuations-to-entropy-production ratio that is of the \emph{order} of one, it is still far from being optimal with respect to the bound Eq.~\eqref{eq:classical_bound_4T}. 
The smaller the colder temperature gets (hence with increasing $T_1-T_2$), the more the ratios approach 1. This is in agreement with the results presented in Fig.~\ref{fig:cooling_close_bound}, where the close-to-optimal values of $\mathcal{R}$ are reached for small cooling powers. By contrast, the ratios decrease with increasing absolute values of the resource fluctuations and the entropy production, which was not the case in the example discussed in Sec.~\ref{sec:approaching}.
Finally, we find a non-differentiable point in the three ratios as function of the temperature difference. This is due to a change of sign in $\Isigma_3$ (from negative to positive), and happens at a temperature close to  the temperature at which the global maximum of $J_\text{cool}$ changes.

\subsubsection{N-demon for power production}

We now analyze the N-demon operating as an engine, with $T_2=T_1$ and $\mu_1\neq \mu_2$ producing electrical power in the working substance, by driving a current against the potential bias. While many of the results that we find are analogous to the results of the refrigerator setup, we here point out relevant differences.

We first of all remark that the parameter regimes in which the demon conditions are fulfilled while useful work is performed in the working substance, are more restricted with respect to the refrigerator setup of the previous subsection, leading to the different shape of the density plot in Fig.~\ref{fig:engine_config1}. This is not a general statement---see also Ref.~\cite{Hajiloo_2020}---but is a consequence of the complex interplay among the system parameters fixed by the demon conditions, which are influenced by the other freely chosen parameters in the specific setting. Also, the specific implementation chosen here leads to a relatively low output power.
We again find two different operational points where the output power has a local maximum, shown as bright spots in Fig.~\ref{fig:engine_config1}(a). 
We note that the sensitivity of \rfs and entropy production, but also of their ratio, on the potential bias is less pronounced than the temperature-bias dependent case discussed for the refrigerator before.

An interesting observation can be made here concerning the behavior of the output power as compared to the resource fluctuations, the entropy production and their ratios.
While in the case discussed in Sec.~\ref{sec:approaching} both ratio and resource fluctuations increase with the increase of the maximum cooling power, in the four-terminal refrigerator realization of Fig.~\ref{fig:refrigerator_config1} the ratio increases whereas the resource fluctuation decrease with the increase of $\max{(J_\text{cool})}$.
Here, by contrast, we have both features: for positive biases the ratio increases while the resource fluctuations decrease as the maximum power increases, whereas at negative biases the opposite happens.
These scenarios are all in line with the bounds on entropy production with respect to resource fluctuations developed in this paper, but they clearly show that the efficiency (namely the ratio between useful output power in the working substance and entropy production in the resource) of the setup does not correlate with the resource fluctuations. The open question under which conditions the \textit{output} (or the efficiency~\cite{Hajiloo_2020}) can be optimized given a certain amount of fluctuations is a motivation for future studies relating our findings to generalized thermodynamics uncertainty relations, see also Eq.~\eqref{eq:TUR}.

\section{Conclusions and outlook}
In this work, we have analyzed the role of fluctuations in electronic mesoscopic engines powered by nonthermal resources. The key result, Eq.~\eqref{eq:result_intro}, is that whenever a given performance goal in the working substance is set, a minimum amount of particle and local entropy fluctuations in the resource is required. {Conversely, a given amount of resource fluctuations sets an upper bound on the useful work that can be done in the system. Interestingly, we find that this bound is set by the \emph{classical}} part of the fluctuations only: When quantum fluctuations are present the noise in the resource part of the system can be lowered, as shown by Eq.~\eqref{eq:fullbound_inf}. 
Based on the inequality~\eqref{eq:result_intro}, we have introduced a notion of \rfs which we have used to characterize different implementations of N-demons.
Considering both the case when the resource is provided by a single contact with a nonthermal occupation probability (Sec.~\ref{sec:approaching}) and the case of a multi-terminal resource effectively implementing a nonthermal input distribution (Sec.~\ref{sec:mixing} and~\ref{sec:analysis}), we have shown that the bound imposed by the derived inequalities can be closely approached even when the cooling power or the produced electrical power are finite and the demon conditions (imposing no average particle and energy exchange between resource region and working substance) are fulfilled.
The bounds developed in this work provide a general statement on the requirement on fluctuations of an engine resource.

How these predictions can be more closely connected to the currently broadly studied thermodynamic uncertainty relations, as alluded to in Secs.~\ref{sec:generalConstraints} and \ref{sec:composite}, will be an interesting topic of future research. This is expected to give even further insights into the operation of nonthermal machines.

\acknowledgements
We thank Henning Kirchberg for useful comments on the manuscript. Funding by the Knut and Alice Wallenberg foundation via the fellowships program (M.A., L.T., J.E., and J.S.), by the Spanish Ministerio de Ciencia e Innovaci\'on via grant No. PID2019-110125GB-I00 and PID2022-142911NB-I00, and through the ``Mar\'{i}a de Maeztu'' Programme for Units of Excellence in R{\&}D CEX2018-000805-M (R.S.), by French ANR projects TQT (ANR-20-CE30-0028) and QuRes (ANR-PRC-CES47-0019) (R.W.), and by the Swedish Vetenskapsr\r{a}det via project number 2018-05061 (J.S.) is gratefully acknowledged.

\appendix

\section{Fluctuating {particle, energy, and entropy currents}}\label{app_entropyoperator}

Using the short-hand notation $\hat{b}_\alpha'\equiv \hat{b}_\alpha(E')$ and dropping the energy arguments, we {first} define {for reference the fluctuating particle- and energy-current operators~\cite{Blanter2000,moskalets-book,Butcher1990Jun},}
\begin{equation}
{\hat{I}^X_\alpha \equiv \int \frac{dEdE'}{h}e^{-\frac{i(E-E')t}{h}} x_\alpha \Big(\hat{b}_\alpha'^\dagger\hat{b}_\alpha - \hat{a}_\alpha'^\dagger\hat{a}_\alpha\Big)}
\end{equation}
{where $\hat{a}_\alpha$ and $\hat{b}_\alpha$ are the field operators of electrons flowing out of and into contact $\alpha$, respectively.} {The operators $\hat{b}_\alpha'^\dagger\hat{b}_\alpha - \hat{a}_\alpha'^\dagger\hat{a}_\alpha$ describe the rate of change of the contact's occupation $f_\alpha$, while providing a probabilistic nature through the random tunnelings of electrons.} {The factor $x_\alpha$, as defined in the main text, determines the transported quantity at every energy, namely $e$ or $E$ for particle and energy current.}

{Next, we here also introduce the} entropy flow into contact $\alpha$ as
\begin{equation}\label{app:eq:fluctuating-entropy-production}
\begin{split}
    \hat{I}^\Sigma_\alpha & \equiv  -\kB \int \frac{dEdE'}{h}e^{-\frac{i(E-E')t}{h}} \Big[ \Big(\hat{b}_\alpha'^\dagger\hat{b}_\alpha - \hat{a}_\alpha'^\dagger\hat{a}_\alpha\Big)\log f_\alpha +\\
    &\qquad\qquad+\Big(\hat{b}_\alpha'\hat{b}_\alpha^\dagger - \hat{a}_\alpha'\hat{a}_\alpha^\dagger\Big)\log(1-f_\alpha) \Big].
\end{split}
\end{equation}
The function $f_\alpha$ describes the occupation number in contact $\alpha$, i.e. $\langle \hat{a}_\alpha'^\dagger \hat{a}_\alpha\rangle = \delta(E-E')f_\alpha$.
Crucially, transport does not affect the occupation of the contact because the latter is considered to be a large bath or, alternatively, its distribution is kept constant by an external agent.
{The formulation of Eq.~\eqref{app:eq:fluctuating-entropy-production} relies on concepts borrowed from information theory. In particular, the Shannon information (or surprisal) associated with an event occurring with probability $p$ is given by $-\log p$. From this, one obtains the Shannon entropy, which is the expected value of the information. For the electronic system we are considering here, a channel originating from contact $\alpha$ can be either occupied (with probability $f_\alpha$) or empty, with probability $(1-f_\alpha)$. So, this defines the two possible outcomes of the event ``electron being present in the channel'', yielding information $-\log(f_\alpha)$ and $-\log(1-f_\alpha)$, respectively. Correspondingly, one can define the information \emph{change} associated with these outcomes as in Eq.~\eqref{app:eq:fluctuating-entropy-production}, where the information is combined with the rate of change of the reservoir occupation, described via the operator difference $\hat{b}^\dagger_\alpha \hat{b}_\alpha-\hat{a}^\dagger_\alpha \hat{a}_\alpha$. Equipped with these definitions, one can find the entropy change as the expectation value of the information change, $\Isigma_\alpha=\braket{\hat{I}^\Sigma_\alpha}$, reproducing the expression for the average entropy in the main text, namely Eq.~\eqref{eq:currents}, with the prescription~\eqref{eq:entropyfactor}. Similarly, one can obtain the entropy fluctuations of Eq.~\eqref{eq:noise-general}.}

{Importantly, this formulation of the information flow relies on the fact that the particles considered here are  fermions. Hence, the only considered occupations are zero or one. }

\section{Current fluctuations in an engineered N-demon}\label{app:current-fluctuations}

We consider a nonthermal resource consisting of a set of thermal contacts connected through a coherent conductor,  see Fig.~\ref{fig:setup_4T}. However, instead of considering the contributing contacts separately---as done in the main text in Eq.~\eqref{eq:sum_demon_fluctuations}---we here consider the \textit{total} (entropy) currents flowing out of the resource region as well as their fluctuations. We show in this appendix that the bound between entropy currents and their fluctuations is 
modified due to cross-correlations between currents into the different resource contacts.

\subsection{Generic multi-terminal setup}

We consider the classical fluctuations of a generic current $\hat{I}^X_\text{N} = \sum_{\alpha\in\mathfrak{N}} \hat{I}^X_\alpha$, where $\mathfrak{N}$ is a collection of contacts, defining the resource part of the system. The fluctuations of interest in this multi-terminal resource are given by the combination of correlators $S_\text{NN,cl}^{\mathrm{multi},X} = \sum_{\alpha\beta\in\mathfrak{N}} S_{\alpha\beta,\text{cl}}^X$. From Eq~\eqref{eq:noise-general}, we write it as
\begin{widetext}
\begin{equation}\label{app:eq:demon-current-fluctuations}
    \begin{split}
        S_\text{NN,cl}^{\mathrm{multi},X} &= \frac{2}{h}\int dE \sum_{\alpha\beta\in\mathfrak{N}} x_\alpha x_\beta \Big[-F_{\alpha\alpha}|s_{\beta\alpha}|^2 - F_{\beta\beta}|s_{\beta\alpha}|^2
        +\delta_{\alpha\beta}\sum_\gamma |s_{\alpha\gamma}|^2(F_{\alpha\gamma} + F_{\gamma\alpha})\Big]\\
        &= \frac{2}{h}\int dE \left[\sum_{\substack{\alpha\beta\in\mathfrak{N},\\\alpha\neq\beta}} x_\alpha x_\beta\Big(-F_{\alpha\alpha} |s_{\beta\alpha}|^2 - F_{\beta\beta}|s_{\alpha\beta}|^2\Big)
        +\sum_{\alpha\in\mathfrak{N}}x_\alpha^2 \sum_{\gamma\neq\alpha}|s_{\alpha\gamma}|^2(F_{\alpha\gamma}+ F_{\gamma\alpha})\right],
    \end{split}
\end{equation}
\end{widetext}
where, as in the main text, $x_\alpha$ is the weight associated to the current, specifically, $x_\alpha=1$ for the particle current and $x_\alpha = k_\text{B}\sigma_\alpha$ for the entropy current.
Note that the cross-correlators, $S^X_{\alpha\beta,\mathrm{cl}}$ with $\alpha\neq \beta$ that give rise to the first line in Eq.~\eqref{app:eq:demon-current-fluctuations}, can be either positive or negative depending on the sign of $x_\alpha x_\beta$.
Furthermore, the last line in Eq.~\eqref{app:eq:demon-current-fluctuations} has the same structure as Eq.~\eqref{eq:noise_classical}. Therefore, we can use the inequalities~\eqref{eq:inequalities} to find the following bound
\begin{widetext}
\begin{equation}\label{eq:engineered_bounds}
\begin{split}    
    S_\text{NN,cl}^{\mathrm{multi},\Sigma} + \frac{k_\text{B}^2}{4}S_\text{NN,cl}^{\mathrm{multi}} &\geq \frac{2k_\text{B}^2}{h}\int dE \Bigg\{\sum_{\alpha\in\mathfrak{N}}|\sigma_\alpha| \sum_{\gamma\neq\alpha}|s_{\alpha\gamma}|^2|f_\alpha-f_\gamma|
        +\sum_{\substack{\alpha\beta\in\mathfrak{N},\\\alpha\neq\beta}} \left(\sigma_\alpha \sigma_\beta+\frac14\right)\left(-F_{\alpha\alpha} |s_{\beta\alpha}|^2 - F_{\beta\beta}|s_{\alpha\beta}|^2\right)\Bigg\}\\
        &\geq 2k_\text{B}\sum_{\alpha\in\mathfrak{N}} |\Isigma_\alpha|
        +\frac{2k_\text{B}^2}{h}\int dE \sum_{\substack{\alpha\beta\in\mathfrak{N},\\\alpha\neq\beta}} \left(\sigma_\alpha \sigma_\beta+\frac14\right)\left(-F_{\alpha\alpha} |s_{\beta\alpha}|^2 - F_{\beta\beta}|s_{\alpha\beta}|^2\right).
\end{split}
\end{equation}
\end{widetext}
Note that the second term on the right hand side of this inequality, adding up to the sum of absolute values of entropy productions in the separate contacts, can be negative. This means that the bound on the fluctuations in the total current from the engineered nonthermal resource is less restrictive and the fluctuations can be smaller compared to the sum in Eq.~\eqref{eq:classical_bound_3T}. Note however, that for the example treated in Sec.~\ref{sec:analysis} (see Figs.~\ref{fig:refrigerator_config1} and \ref{fig:engine_config1}), the fluctuations on the left hand side of Eq.~\eqref{eq:engineered_bounds} are larger than the entropy production in the right hand side of 
Eq.~\eqref{eq:sum_demon_fluctuations}.
\subsection{Linear response of a four-terminal system}\label{app_linear_response}
We show here that, within linear response, the particle- and energy-current fluctuations of a single nonthermal terminal, see Fig.~\ref{fig:setup_3T}, and the noise of the total particle and energy currents from an engineered nonthermal two-terminal resource, see Fig.~\ref{fig:setup_4T}, are identical. 
We therefore consider the 4-terminal setup and assume that $(\mu_\alpha-\mu_0)/(k_\text{B}T_0)\ll 1$, $(T_\alpha-T_0)/T_0\ll 1$, for some reference temperature $T_0$ and chemical potential $\mu_0\equiv 0$. 
Then, we can expand all currents up to linear order in the affinities $\mathcal{A}_\alpha^\mu=(\mu_\alpha-\mu_0)/(k_\text{B}T_0)$, $\mathcal{A}_\alpha^T=(T_\alpha-T_0)/(k_\text{B}T_0^2)$. 

\subsubsection{Demon conditions}\label{sec:linear_currents}
With this and with the definition $\mathcal{A}_{\alpha\beta}^{\mu,T}\equiv\mathcal{A}_{\alpha}^{\mu,T}-\mathcal{A}_{\beta}^{\mu,T}$, the demon conditions read
\begin{align*}
    g_\text{int}^{(0)}\mathcal{A}_{23}^\mu+g_\text{int}^{(1)}\mathcal{A}_{23}^T+h_\text{res}^{(0)}\mathcal{A}_{34}^\mu+h_\text{res}^{(1)} \mathcal{A}_{34}^T&=0\\
    g_\text{int}^{(1)}\mathcal{A}_{23}^\mu+g_\text{int}^{(2)}\mathcal{A}_{23}^T+h_\text{res}^{(1)}\mathcal{A}_{34}^\mu+h_\text{res}^{(2)}\mathcal{A}_{34}^T&=0
\end{align*}
where we have defined the integrals
\begin{align}
    g_i^{(\nu)}&=\int dE\,\tau_i(E)\xi(E)E^\nu\,,\\
    h_i^{(\nu)}&=\int dE\,\tau_i(E)\tau_\text{int}(E)\xi(E)E^\nu\,,
\end{align}
with $\xi(E)=-k_\text{B}T_0(\partial f_0/\partial E)$. Using these equations to fix the values of the chemical potentials $\mu_3$ and $\mu_4$, we find
\begin{align}
    \mu_3&=\mu_2+\frac{\beta}{\alpha}\frac{T_3-T_2}{T_0}+\frac{\gamma}{\alpha}\frac{T_4-T_3}{T_0},\\
    \mu_4&=\mu_2+\frac{\delta}{\alpha}\frac{T_3-T_2}{T_0}+\frac{\zeta}{\alpha}\frac{T_4-T_3}{T_0},
\end{align}
with the coefficients
\begin{align}
    \alpha&=h_\text{res}^{(1)}g_\text{int}^{(0)}-h_\text{res}^{(0)}g_\text{int}^{(1)},\\
    \beta&=h_\text{res}^{(0)}g_\text{int}^{(2)}-h_\text{res}^{(1)}g_\text{int}^{(1)},\\
    \gamma&=h_\text{res}^{(2)}h_\text{res}^{(0)}-[h_\text{res}^{(1)}]^2,\\
    \delta&=g_\text{int}^{(1)}(g_\text{int}^{(1)}-h_\text{res}^{(1)})+g_\text{int}^{(2)}(h_\text{res}^{(0)}-g_\text{int}^{(0)}),\\
    \zeta&=h_\text{res}^{(1)}(g_\text{int}^{(1)}-h_\text{res}^{(1)})+h_\text{res}^{(2)}(h_\text{res}^{(0)}-g_\text{int}^{(0)}).
\end{align}

\subsubsection{Fluctuations}
All noise contributions can be expanded up to linear order, too. Let us start with the 4-terminal setup and expand the noise~\eqref{eq:noise_demon_4T} at linear order in the affinities. We have
\begin{equation}
    f_\alpha(1-f_\alpha)\approx f_0(1-f_0)+\xi(\mathcal{A}_\alpha^\mu+E\mathcal{A}_\alpha^T)(1-2f_0).
\end{equation}
All the terms of the form $(f_\alpha-f_\beta)^2$ are at least quadratic in the affinities. In particular the quantum part of the noise, Eq.~(\ref{eq:noise_quantum}), (or also the shot noise in other ways of dividing noise contributions) at this order vanishes. We are therefore left with
\begin{subequations}
\label{eq:noise_linear}
\begin{align}
        S_\text{NN}^{\text{4T}}&=2r^{(0)}+(\mathcal{A}_2^\mu+\mathcal{A}_3^\mu)p^{(0)}+(\mathcal{A}_2^T+\mathcal{A}_3^T)p^{(1)}\nonumber\\
        &\quad+\mathcal{A}_{43}^\mu q^{(0)}+\mathcal{A}_{43}^T q^{(1)}   \label{eq:chargenoise_linear}\\
        S_\text{NN}^{\text{4T,E}}&=2r^{(2)}+(\mathcal{A}_2^\mu+\mathcal{A}_3^\mu)p^{(2)}+(\mathcal{A}_2^T+\mathcal{A}_3^T)p^{(3)}\nonumber\\
        &\quad+\mathcal{A}_{43}^\mu q^{(2)}+\mathcal{A}_{43}^T q^{(3)}  \label{eq:energynoise_linear}
    \end{align}  
\end{subequations}
where
\begin{align}
    r^{(\nu)}&=\int dE\,\tau_\text{int}(E)f_0(E)[1-f_0(E)]E^\nu,\\
    p^{(\nu)}&=\int dE\,\tau_\text{int}(E)\xi(E)[1-2f_0(E)]E^\nu,\\
    q^{(\nu)}&=\int dE\,\tau_\text{int}(E)\tau_\text{res}(E)\xi(E)[1-2f_0(E)]E^\nu.
\end{align}
If we now consider the result in the 3-terminal setup with an effective nonthermal distribution $f_\text{N}$ given by Eq.~\eqref{eq:fneq_map}, the noise is given by Eq.~\eqref{eq:noise_demon_3T}. It contains the term $f_2(1-f_2)$, just as the 4-terminal expression~\eqref{eq:noise_demon_4T}, and the terms $f_\text{N}(1-f_\text{N})$ and $(f_2-f_\text{N})^2$. It is easy to see that the latter does not contribute at linear order in the affinities, while the remainder gives
\begin{equation}
\begin{split}
    f_\text{N}(1-f_\text{N})&=f_3(1-f_3)+\tau_\text{res}(f_4-f_3)(1-2f_3)\\
    &\quad-\tau_\text{res}^2(f_3-f_4)\\
    &\approx f_0(1-f_0)+\xi(1-2f_0)(\mathcal{A}_3^\mu+E\mathcal{A}_3^T)\\
    &\quad+\tau_\text{res}\xi(1-2f_0)(\mathcal{A}_{43}^\mu+E\mathcal{A}_{43}^T)
\end{split}
\end{equation}
which shows that the noise~\eqref{eq:noise_demon_3T} for the 3-terminal N-demon becomes identical to the expression given in Eq.~\eqref{eq:noise_linear} in linear response.

\section{Analytical expressions for energy noise with QPC transmissions}\label{app_analytical}

Useful analytical expressions for charge and energy currents in the presence of sharp step-function transmissions can be found in Appendix~A of Ref.~\cite{Hajiloo_2020}. Here we present some analytical expressions for the noise, in the case when all terminals are described by a thermal distribution, as considered in Sec.~\ref{sec:composite}.
The calculation of the energy noise [see, e.g., Eq.~\eqref{eq:noise_demon_4T}] involves integrals of the form
\begin{equation}
    \mathcal{J}_\alpha=\int_{E_0}^{+\infty} dE E^2 f_\alpha(E)[1-f_\alpha(E)]
\end{equation}
where $f_\alpha(E)$ is a thermal distribution and $E_0$ is the onset energy of a sharp step function transmission $\tau_\text{QPC}=\Theta(E-E_0)$. This integral can be evaluated analytically, yielding
\begin{equation}
\begin{split}
    \mathcal{J}_\alpha&=\mathcal{J}_\alpha^\text{open}+(\kB T_\alpha)^3\big[2x_\alpha\log(1+e^{x_\alpha-y_\alpha})\\
    &\quad+2\mathrm{Li}_2(-e^{x_\alpha-y_\alpha})-x_\alpha^2(1+e^{y_\alpha-x_\alpha})^{-1}\big],
\end{split}
\end{equation}
where $\text{Li}_2$ is the dilogarithm function, $x_\alpha=E_0/(\kB T_\alpha)$, $y_\alpha=\mu_\alpha/(\kB T_\alpha)$, and
\begin{equation}
    \mathcal{J}_\alpha^\text{open}=(\kB T_\alpha)^3\left[\frac{\pi^2}{3}+y_{\alpha}^2\right]
\end{equation}
is the result for an open channel with $E_0\to-\infty$.
Integrals involving factors of the form $(f_\alpha-f_\beta)^2$ do not have a closed form when the Fermi functions have different temperatures and we resort to a numerical integration to evaluate them.

\bibliography{Refs.bib}

\end{document}